\documentclass[11pt]{article}

\usepackage{times}
\usepackage{style}
\usepackage{fullpage}

\usepackage{hyperref}
\hypersetup{
     colorlinks   = true,
     citecolor    = gray,
     linkcolor    = blue
}

\newcommand{\wcost}{\omega}
\newcommand{\memsize}{M}
\newcommand{\ourmodel}{$(\memsize,\wcost)$-ARAM}
\newcommand{\ourmodelfull}{$(\memsize,\wcost)$-Asymmetric RAM (ARAM)}

\newcommand{\smallmem}{small-memory}
\newcommand{\largemem}{large-memory}

\newcommand{\ourht}{$k$-level hash table}

\newcommand{\rotating}{rotating arrays}

\newcommand{\cm}{read transfer}
\newcommand{\wrt}{write transfer}
\newcommand{\rwt}{read and write transfer}

\newcommand{\classic}{\textbf{Classic}}
\newcommand{\static}{\textbf{Static}}
\newcommand{\splitpool}{\textbf{SplitPool}}

\newcommand{\SPLIT}{\textsc{Split}}
\newcommand{\JOIN}{\textsc{Join}}

\newcommand{\whp}{\emph{whp}}

\title{Algorithmic Building Blocks for Asymmetric Memories}

\author{Yan Gu\\Carnegie Mellon University\\yan.gu@cs.cmu.edu \and Yihan Sun\\Carnegie Mellon University\\yihans@cs.cmu.edu \and Guy E.\ Blelloch\\Carnegie Mellon University\\guyb@cs.cmu.edu}

\date{}

\begin{document}

\maketitle

\begin{abstract}
The future of main memory appears to lie in the direction of new non-volatile memory
technologies that provide strong capacity-to-performance ratios, but
have write operations that are much more expensive than reads in terms
of energy, bandwidth, and latency.
This asymmetry can have a significant effect on algorithm design, and in many cases it is possible to reduce writes at the cost of reads.
In this paper we study which algorithmic techniques are useful in designing practical write-efficient algorithms.
We focus on several fundamental algorithmic building blocks including unordered set/map implemented using hash tables, ordered set/map implemented using various binary search trees, comparison sort, and graph traversal algorithms including breadth-first search and Dijkstra's algorithm.
We introduce new algorithms and implementations that can reduce writes, and analyze the performance experimentally using a software simulator.
Finally we summarize interesting lessons and directions in designing write-efficient algorithms.
\end{abstract}

\section{Introduction}

The future of main memory appears to lie in the non-volatile
memory technologies that promise persistence, significantly lower energy costs, and higher
density than the DRAM technology used in today's main
memories~\cite{hp-nvm15, intel-nvm15, Meena14, Yole13}.
Despite the advantages, a key property of such memory technologies, however, is their asymmetric read-write
costs: compared to reads, writes can be much more expensive in terms of latency, bandwidth, and energy.
Because bits are stored in these
technologies as at rest ``states'' of the given material that can be
quickly read but require physical change to update, this asymmetry
appears fundamental.
This motivates the need for \emph{write-efficient} algorithms that
largely reduce the number of writes compared to existing algorithms.

In the related work section, we review the literature on studying this read-write asymmetry on NAND
Flash chips~\cite{BT06, Eppstein14, Gal05, ParkS09} and algorithms targeting database operators~\cite{Chen11,Viglas12, Viglas14}.
These works provide novel aspects on rethinking algorithm design. 
However, most of the papers either treat NVMs as external memories, or are based on hardware simulators for existing architecture, which may have many concerns that we will further discuss in the related work section.

Blelloch et al.~\cite{BBFGGMS16,BFGGS15,blelloch2016parallel}
formally defined and analyzed several sequential and parallel computation models with good caching and scheduling guarantees.
The models abstract such asymmetry between reads and writes, and can be used to analyze algorithms on future memory.
The basic model, which is the Asymmetric RAM (ARAM), extends the well-known external-memory model~\cite{AggarwalV88} and parameterizes the asymmetry using $\wcost$, which corresponds to the cost of a write relative to a read to the non-volatile main memory.
The cost of an algorithm on the ARAM, the \textbf{asymmetric I/O cost}, is the number of write transfers to the main memory multiplied by $\wcost$, plus the number of read transfers.
This model captures different system consideration (latency, bandwidth, or energy) by simply plugging in different values of $\wcost$,
and also allows algorithms to be analyzed theoretically.
Based on this idea, many interesting algorithms (and lower bounds) are designed and analyzed by various recent papers~\cite{BBFGGMS16,bbfgg2017implicit,BFGGS15,blelloch2016efficient,blelloch2018geometry,Jacob17}.

Unfortunately, all of the analyses of such write-efficient algorithms are asymptotic, showing the upper and lower bounds on the complexity of these problems.
Also, to prove the bounds, the theoretical models simplify the real architecture (e.g.,\ without considering blocking of cache-lines or cache policies).
It still remains unknown what the performance of these algorithms are in practice.
In this paper, our goal is to show such performance on a number of fundamental algorithmic building blocks.
We believe the lessons in designing and implementing them are useful for our community to use new memory in the future.

\myparagraph{Contribution of this paper.}

In this work, our goal is to bridge the gap between theory and practice.
We try to study and understand which algorithmic techniques are useful in designing practical write-efficient algorithms.
As the first paper of this kind, we focus on several of the most commonly-seen algorithmic building blocks in modern programming: unordered set/map implemented using \textbf{hash tables},
set/map implemented using \textbf{balanced binary search trees},
\textbf{comparison sort}, and graph traversal algorithms: \textbf{breadth-first search} for unweighted graphs and \textbf{Dijkstra's algorithm} for weighted graphs\footnote{The algorithms that are not I/O-bounded or use much fewer writes than reads are not discussed in this paper (e.g., matrix multiplication and other algorithms with a similar computation pattern~\cite{BFGGS15}).}.

Unfortunately, no non-volatile main memory is currently available,
making it impossible to get real timings.  Furthermore, details about
latency and other parameters of the memory and how they will be
incorporated into the architecture are also not available.  This makes
detailed cycle-level simulation (e.g., PTLsim~\cite{ptlsim},
MARSSx86~\cite{marss86} or ZSim~\cite{sanchez2013zsim}) of questionable utility.
However, it is quite feasible to count the number of reads and write to main memory
while simulating a variety of cache configurations.
For I/O-bounded algorithms, these numbers can
be used as reasonable proxies for both running time (especially when implemented in parallel) and energy consumption.\footnote{\small The energy consumption of main memory is a key concern since it costs 25-50\% energy on data centers and servers~\cite{lefurgy2003energy,malladi2012rethinking,leverich2014reconciling}.}
Moreover, conclusions drawn from these numbers can likely give insights into
tradeoffs between reads and writes among different algorithms.

For these reasons, we propose a
framework based on a software simulator that can efficiently and
precisely measure the number of \rwt{s} of an algorithm using
different caching policies.  We also consider variants in caching
policies that might lead to improvements when read and write are not
the same.


We also note that designing write-efficient algorithms falls in a high dimensional parameter space since the asymmetries on latency, bandwidth, and energy consumption between reads and writes are different.   Here we abstract this as a single value
$\wcost{}$.
This value together with the cache size $\memsize$ and cache-line size $B$ (set to be $64$ bytes in this paper) form the parameter space of an algorithm.

Our framework provides a simple, clean and hardware-independent method to analyze and experiment the performance on the asymmetric memory.
We investigate the algorithmic techniques and learn lessons from the experiments that generally apply for a reasonably large parameter space of $\wcost$, $M$ and $B$.
This framework also allows monitoring, reasoning and debugging the code easily, so it can remain useful even after the new hardware is available.

\medskip

\ifx\conferenceversion\undefined
With the framework, we design, implement and discuss many different algorithms and data structures and their write-efficient implementations.
Although some of the implementations are standard, like quicksort and the classic hash tables, many others, including the \ourht{s}, sample sort\hide{, red-black trees} and phased Dijkstra, require careful algorithmic design, analysis, and coding.
Under our cost measure, which is the asymmetric I/O cost, we show better approaches on all problems we study in this paper, compared to the most basic and commonly-used ones on symmetric memories.
We understand that there are more advanced versions of the algorithms and data structures discussed in this paper on some specific applications, and how to implement them write-efficiently is an interesting topic for future work.


\else
With the framework, we design, implement and discuss many algorithms and data structures and their write-efficient implementations.
Although some of the implementations are standard, like quicksort and hash tables, many others, including \ourht{s}, sample sort and phased Dijkstra, require careful algorithmic design, analysis, and coding.
Under our measurement which is the asymmetric I/O cost and compared to the most commonly-used ones on symmetric memories, we provide better alternatives to all problems we studied in this paper.
\fi

With the algorithms and their experimental results, we draw many interesting algorithmic strategies and guidance in designing write-efficient algorithms.
A common theme is to trade (more) reads for (fewer) writes (apparently it is hard to directly decrease the writes since this can improve the performance on symmetric memory as well and should have been investigated already).
Some interesting lessons we learned and can be valuable to share are listed as follows, which can suggest some potential directions to design and engineer write-efficient algorithms in the future.
\begin{compenumerate}
  \item Indirect addressing is less problematic.  In the classic setting, indirect addressing should be avoided if possible, since each addressing can be a random access to the memory.  However, when writes are expensive, moving the entire data is costly, while indirect addressing only modifies the pointers (at the cost of a possible random access per lookup).
      \hide{
  \item Balancing (like in BSTs) is less critical.  A more balanced tree has smaller average path length (i.e., \ reads) but requires more effort to maintain the structure (i.e., \ writes).  Less balancing should be involved if there are frequent updates.}
  \item Multiple candidate positions for a single entry in a data structure can help.
      It can be a good option to use more reads per lookup but apply less frequent data movements, when the size of a data structure changes significantly.
      This is a common strategy we have applied in this paper to provide an algorithmic tradeoff between reads and writes.
  \item It is usually worth to investigate existing algorithms that move or modify the data less.  These algorithms can be less efficient in the symmetric setting due to various reasons (e.g., more random accesses, less balanced), but the property that they use fewer writes can be useful in the asymmetric setting (like samplesort vs.\ quicksort, treap vs.\ AVL or red-black tree).
  \item {In-cache data structures} should draw more attention\footnote{\small Similar ideas appear in many existing models already, like the external-memory model or the streaming model.  However, the motivations are different: these models restrict the amount of space that can be used in the computation, while in our case the data structures are used to reduce the writes to the asymmetric main memory without including too many extra reads.}.  Since the data structures are kept in the cache (or small symmetric memory), the algorithm requires significantly less writes to the large asymmetric memory, although may require extra reads to compensate for less information we can keep within the data structure.
      In this paper, we discuss Dijkstra's algorithm on shortest-paths as an example, and such idea can also be applied to computing minimum spanning tree, sorting, and many other problems.
\end{compenumerate}

\hide{
To design write-efficient algorithms in the symmetric setting, we suggest that the following directions can be investigated.
\begin{compenumerate}
  \item Investigating existing algorithms that require less effort to maintain the states or elements in the algorithm.  These algorithms can be less efficient in the symmetric setting, but their special properties that require fewer updates can be useful in the asymmetric setting (like samplesort vs.\ quicksort, treap vs.\ AVL or red-black tree).
  \item Providing algorithmic tradeoff between reads and writes.  It is usually worthwhile to reduce the write bottleneck by some algorithmic improvement that leads to more reads.   For example in this paper, we show the \ourht{} that requires more reads for a query but significantly reduces the number of writes in resizing compared to standard hash tables.  Our \ourht{}, as well as the graph traversal algorithms using it, has a consistently better overall performance.
  \item Designing algorithms using in-cache data structures.  In this paper we discussed Phased Dijkstra, which is a variant of Dijkstra's algorithm that fit all computations in the cache (i.e.\ requiring no writes to the main memory).  The compromise is that the algorithm may require multiple phases to finish, but in most cases this is worthwhile for the reduced writes.  Such an idea can also be applied to computing minimum spanning tree, sorting, and many other problems.
\end{compenumerate}
}
\hide{
We understand that many more advanced algorithms, data structures and implementations are proposed on the fundamental problems we studied in this paper, on different settings (e.g.\ multi-core, many-core, distributed), utilizing new processors and memory technologies, or focusing on specific input distributions to improve the performance.
Although we are unable to implement and compare all of them in this paper, we believe that the general strategies we mentioned above are also the directions to derive the write-efficient versions of these implementations.
}

\section{Related Work}

There exist a rich literature to show the read-write asymmetry on the new memories~\cite{Akel11, Athanassoulis12, BFGGS15, blelloch2016efficient,
  carson2016write, ChoL09, Dong09, Dong08, HuZXTGS14,
  ibm-pcm14b, Kim14, LeeIMB09, malicevic2015exploiting, Qureshi12, Xu11, yang:iscas07,
  ZhouZYZ09, ZWT13}.
Regarding adapting softwares for such read-write asymmetry,
some work has studied the system aspect.
For example, there exist many papers on how to balance the writes across the chip to avoid uneven wear-out of locations
in the context of NAND Flash chips~\cite{BT06, Eppstein14, Gal05, ParkS09}.

More closely-related papers targeting database operators:
Chen et al.~\cite{Chen11} and Viglas~\cite{Viglas12, Viglas14}
presented several interesting and write-efficient sequential algorithms for searching, hash joins and sorting.
These are the early and inspirational attempts to design algorithms with fewer writes.
Instead of formally proposing new computational models and analyzing the asymptotic cost,
they mainly show the performance by the experiment results assuming external memories rather than main memories, or on the cycle-based simulators for existing architecture.
For the latter case however, the prototypes of the new memories are still under development, and yet nobody actually knows the exact parameters of the new memories, or how they are incorporated into the actual architecture which is required for the setup of the cycle-based simulator.
As far as we know, there is no available cycle-based simulator at the present time for the new memories.
In the meantime, the asymmetries on latency, bandwidth, and energy consumption between reads and writes are different, and any of these constraints can be the bottleneck of an algorithm.
Hence, designing algorithms on asymmetric memory are in a multiple-dimension parameter space, rather than just recording the running time from a simulator.
Therefore, it is essential to develop theoretical models and tools that accounts for, and abstract this asymmetry and use them to analyze algorithms on future memory.

Blelloch et al.~\cite{BBFGGMS16,BFGGS15,blelloch2016parallel} formally defined several sequential and parallel computation models that take
asymmetric read-write costs into account.
Based on the computational models, many interesting algorithms (and lower bounds) are designed and analyzed in both sequential and parallel settings, which includes sorting, permuting, matrix multiplication, FFT, list/tree contraction, BFS/DFS and other graph algorithms, and many computational geometry and dynamic programming problems~\cite{BBFGGMS16,bbfgg2017implicit,BFGGS15,blelloch2016efficient,blelloch2018geometry,Jacob17}.
Carson et al.~\cite{carson2016write} also presented write-efficient sequential
algorithms for a similar model, as well as write-efficient parallel
algorithms (and lower bounds) on a distributed memory model with
asymmetric read-write costs, focusing on linear algebra problems and
direct N-body methods.
Although many problems under the asymmetric setting have been studied, all the analyses are asymptotic and only show the upper and lower bounds on the complexity of these problems.

\hide{  For example, most existing libraries implement binary search trees with red-black trees or AVL trees due to their nearly optimal average tree depth (to the root), shown in Table~\ref{tab:ordered}.  However, we show that treap will be more preferable in the asymmetric setting due to its simplicity for balancing maintenance.  Similarly, for comparison sort, most existing libraries are based on quicksort (or its variants), we show that for larger entry size (greater or equal to 16 bytes), BST sort is more ideal, since once an element is inserted into the search tree, it requires no further movement during the whole process.  For general cases, samplesort always outperforms quicksort since the local search tree structure within the cache largely minimizes data movements.
}

\section{Our Model and Simulator}\label{sec:model}

To start with, we discuss how to measure the performance of algorithms on asymmetric memories.
We begin with the computational model that estimates the cost of an algorithm.
This model requires the numbers of \rwt{s} between the non-volatile memory and the cache, so later we introduce how the numbers of an algorithm can be simulated.
Unlike the existing symmetric memories, a simple cache policy like LRU does not work on some asymmetric settings.
Thus in Section~\ref{sec:policy} we briefly summarize the solutions to fix it, and then the cache simulator given in Section~\ref{sec:simulator} captures this number with different cache policies.

\subsection{The Cost Model for Asymmetric Memory}\label{sec:cost-model}

The most commonly-used cost measure of an algorithm is the time complexity based on the RAM model, which is the overall number of instructions and memory accesses executed in this algorithm.
Nowadays, since the actual latency of an access to the main memory is at least two orders of magnitudes more expensive than a CPU instruction, the \emph{I/O cost} based on the external-memory model~\cite{AggarwalV88} is widely used to analyze the cost of an I/O-bounded algorithm.
This model assumes a \emph{\smallmem} (cache) of size $\memsize \geq 1$, and a \emph{\largemem} of unbounded size.
Both memories are organized in blocks (cache-lines) of $B$ words.
The CPU can only access the \smallmem{} (with no cost), and it takes unit cost to transfer a single block between the \smallmem{} and the \largemem{}.
This cost measure estimates the running time reasonably well for I/O-bounded algorithms, especially in multi-core parallelism.
An efficient algorithm in practice should achieve optimality in both the time complexity and the I/O cost.

To account for more expensive writes on future memories, here we adopt the idea of an \ourmodelfull{}~\cite{blelloch2016efficient}: similar to the external-memory model, transferring a block from \largemem{} to \smallmem{} takes unit cost; on the other direction, the cost is either 0 if this block is clean and never modified, or $\wcost{}\gg 1$ otherwise.
The \textbf{asymmetric I/O cost $Q$}\footnote{Throughout the paper, we abbreviate it as the \emph{I/O cost}, unless stated otherwise explicitly.} of an algorithm is the overall costs for all memory transfers. Theoretical results on this new model have been studied in~\cite{BBFGGMS16,bbfgg2017implicit,BFGGS15,blelloch2016efficient,blelloch2018geometry,Jacob17}. 

\subsection{Cache Policies}
\label{sec:policy}

Either the classic external-memory model or the new ARAM assumes that we can explicitly manipulate the cache in the algorithm.  This largely simplifies the analysis, and in many cases is provably within a constant factor of a more realistic cache's performance.
For example, the standard least-recent used (LRU) policy is 2-competitive against the optimal offline cache-replacement sequence.

Interestingly, the competitive ratio does not hold in the asymmetric setting.
Consider a cache with $k=M/B$ cache-lines and a memory access pattern that repeatedly writes to $k-1$ cache-lines and read from other $k-1$ cache-lines.
An ideal cache policy will keep all $k-1$ cache-lines associated to writes, so the I/O cost of each round is $k-1$ for $k-1$ read misses.
An LRU policy however causes a cache miss for every single memory access, leading the I/O cost of each round to $\wcost(k-1)+k-1$.
This overhead is proportional to $\wcost$, which can be significant and problematic.

The solution is affected by the architecture, depending on whether software explicitly controls a DRAM buffer or not~\cite{Chen11,condit2009better,lee2009architecting,qureshi2009scalable}.
If so, then the cost measures on the these models are just the costs in practice, but programmers are responsible for managing what to put on the \smallmem{} and guaranteeing correctness.
The other option is to leave the hardware to control the \smallmem{}.
In this case, Blelloch et al.~\cite{BFGGS15} show that if the \smallmem{} is partitioned into two equal-size pools and each of them is maintained using LRU policy, the performance is 3-competitive against the optimal offline cache-replacement sequence (e.g.\,using $3\times$ space and incurring no more than $3\times$ cost).

We consider three different cache policies in this paper.
The \classic{} policy maintains the \smallmem{} as one memory pool and uses the LRU policy for replacement.
The \splitpool{} policy keeps two separate memory pools and each runs the LRU policy.
The \static{} policy allows static allocation in one memory pool, and the unallocated memory space is maintained using the LRU policy. 

\subsection{The Cache Simulator}
\label{sec:simulator}

\ifx\database\undefined
The goal of this paper is to discuss new algorithmic approaches that minimize the asymmetric I/O cost to the main memory on a variety of fundamental data structures and algorithms.
\fi
To capture the number of reads and writes to the main memory, we developed a software simulator that can adapt to different cache policies introduced in Section~\ref{sec:policy}.
The cache simulator is composed of an ordered map that keeps tracks of the time stamp of the last visit to each cache-line in the current cache, and an unordered map that stores the mapping from each cache-line to the corresponding location in the ordered map if this cache-line is currently in the cache.  Interestingly, the implementation of this cache simulator is a natural application of the techniques discussed in this paper.

The cache simulator encapsulates a new structure {\sc Array} that is used in coding algorithms in this paper.
It is like a regular array that can be dynamically allocated and freed, and supports two functions: {\sc Read} and {\sc Write} to a specific location in this array.
The {\sc Array}s are responsible for reporting the memory accesses of the algorithm to the cache simulator, and the cache simulator will update the state of the cache accordingly.
Therefore, coding using the {\sc Array}s is not different from regular programming much.

The memory accesses to loop variables and temporary variables are ignored, as well as the call stack.
This is because the number of such variables is small in all of the algorithms in this paper (usually no more than 10).  
Meanwhile, the call stack of all algorithms in this paper has size $O(\log n)$.  The overall amount of uncaptured space is orders of magnitudes smaller than the amount of fast memory in our experiments.

The cache consists of one or two memory pools, depending on different cache policies discussed in Section~\ref{sec:policy}.  We will explicitly indicate the cache policy used in each of our experiments.
The cache simulator maintains two counters in each memory pool: the number of \textbf{\cm{}s}, and the number of \textbf{\wrt{}s}.
When testing each algorithm on a specific input instance, the cache is emptied at the beginning and flushed at the end. A read or write is free if the location is already in the cache; otherwise the corresponding cache-line is loaded, the counter of \cm{} increments by 1, and the least-recently-used cache-line in this pool is evicted.
Also, a write will mark the dirty-bit of the cache-line to be \texttt{true}.
When evicting a dirty cache-line, the counter of \wrt{} increments by 1.
Notice that memory reads can cause \wrt{}s, and memory writes can lead to \cm{}s.

When simulating the \classic{} policy (i.e., the standard one), we also verified our simulated results to ZSim (cycle-level simulator for current architecture), and the numbers always differ by no more than 10\% when the parameters are set correctly.

\section{Sets and Maps}

Sets and maps are two of the most commonly-used data types in modern programming.
Most programming languages either have them built in as basic types (e.g.\ python) or
supply them as standard libraries (C++, C\#, Java, Scala, Haskell, ML).
In this section we discuss efficient implementations of unordered sets and maps implemented using hash tables, and due to the page limit, in ordered sets and maps implemented using balanced binary search trees are introduced in the full version of this paper.

\hide{
These data types and the data structures to implement them are used extensively.
Not only many other data types can be built based on them,
just in the content of this paper, unordered sets and maps are used in BFS in Section~\ref{sec:bfs}, Dijkstra's algorithm in Section~\ref{sec:dijk}, and the cache simulator, while ordered sets and maps are used in Dijkstra's algorithm in Section~\ref{sec:dijk} and the cache simulator.
For example, a priority queue can be implemented based on an efficient implementation of ordered sets, especially in parallel.
Range queries can be built based on binary search trees, so that the data structure like range trees, interval trees, etc., can be efficiently implemented using an appropriate implementation of augmented ordered maps using BSTs~\cite{sun2016pam}. \yan{revise this paragraph.}}

\subsection{Unordered Sets and Maps}\label{sec:unordered}

Our implementation of unordered sets and maps is based on hash tables that support \textbf{lookup}, \textbf{insertion}, and \textbf{deletion}. 
The hash tables discussed in this section use open addressing and linear probing, since the goal of the data structure is to try to minimize the I/O cost focusing on smaller entries (accessing and reading larger entries are costly anyway so different hash-table implementations make minor differences).
For simplicity, we assume no duplicate keys, and it is straightforward to handle the duplicates with minor modifications.
In this setting, each operation of the hash table reads a small number of cache-lines, and an insertion or deletion will modify exactly one cache-line that contains the location of the key and will be eventually written back to the \largemem{}.

The challenge emerges when the set size changes dynamically.
For an efficient implementation, we hope the overall size of the hash table to be neither too large nor too small.
If the load factor passes 80\%, linear probing's performance drastically degrades.
On the other hand, we want the hash table size to be reasonably small to better utilize the \smallmem{} (cache), since each cache-line holds more entries in this case.
In practice, some implementations keep the load factor up- and lower-bounded by some constant.
For example, a typical implementation keeps the occupancy of the hash table between 1/8 and 1/2, and the size doubles or shrinks by half if the number of entries exceeds this range.
Such resizing reinserts $p$ entries before at least $p/2$ insertions and deletions (where $p$ is the set/map size).
When reads and writes have approximately the same cost, the extra cost for such resizing is small compared to the query and update costs (e.g., the queries read from lots of memory locations).
In the asymmetric setting however, the reads cost much less, but the extra writes in resizing can be significant: the resizing can incur at most twice ($p/(p/2)=2$) the writes compared to the initial insertions ($3\times$ writes in total).
Hence, our goal is to discuss an alternative approach that optimizes such extra writes.

\subsubsection{The k-level Hash Table}
\label{sec:hashtable}

Instead of keeping one hash table, our main idea is to maintain a small number $k$ of hash tables simultaneously, where $k$ is a pre-determined parameter.
In particular, the \ourht{} $\mb{HashTable}$ is initialized with $k$ arrays $\mb{HashTable}_{1,\cdots, k}$ with size $2^{c'+i}$ for $1\le i\le k$ (or smaller in specific applications) and a constant $c'$. In practice we set $c'$ to be $5$.

For insertions, when the overall load factor exceeds some threshold $r$, we allocate a new chunk of memory with the double size of the largest current array, and the smallest hash table is discarded after all elements in it have been reinserted back.
Similarly for deletions, if the occupancy of the hash tables drops below a threshold $l$, a small array with half of the size of the current smallest hash table is allocated, and the largest table is freed after the entries in it being reinserted.
For instance, a valid \ourht{} may contain two arrays of size $2^{15}=32768$ and $2^{16}=65536$, when $k=2$ and $30000$ entries in the current configuration.
The pseudocode of the \ourht{} is given in Algorithm~\ref{algo:hash-table}.
The occupancy range $0<l<r<1$ indicates when the resizing happens (a valid set of parameters can be $1/8$ and $1/2$).
A classic implementation can be viewed as the special case of the \ourht{} when $k=1$.

\begin{algorithm}[t]
\small
\caption{The \ourht{}}
\label{algo:hash-table}
\fontsize{10pt}{10pt}\selectfont
\KwIn{Parameter $k$, occupancy range $l$ and $r$}
    \vspace{0.5em}

    \myfunc{\textsc{Lookup($x$)}} {
        \For {\upshape $i \gets 1$ to $k$} {
            $p \gets \mb{HashTable}_i.\textsc{Lookup}(x)$\\
            \lIf {$p \ne \mb{null}$} { \Return{$(i,p)$} }
        }
        \Return{\mb{null}}
    }
\medskip
    \myfunc{\upshape\textsc{Insert($x$)} // $x$ is not in $\mb{HashTable}$} {
        \For {\upshape $i \gets 1$ to $k$} {
            \If {\upshape $\mb{HashTable}_i.\mb{occupancy} < r$} {
                $\mb{HashTable}_i.\textsc{Insert}(x)$\\
                \Return
            }
        }
        Allocate $\mb{HashTable}_{k+1}$ of size $2\cdot \mb{HashTable}_{k}.\mb{size}$\\
        Relabel the hash tables with indices from $0$ to $k$\\
        \ForEach {$y \in \mb{HashTable}_{0}$} {
            \textsc{Insert($y$)}
        }
        Free $\mb{HashTable}_{0}$
    }
\medskip
    \myfunc{\upshape\textsc{Delete($x$; $i$, $p$)} // $x$ is located $p$-th in $\mb{HashTable}_{i}$} {
        $\mb{HashTable}_i.\textsc{Delete}(x,p)$\\
        \If {\upshape Overall occupancy is less than $l$ (and $\mb{HashTable}_{1}.\mb{size}>1$)} {
            Allocate $\mb{HashTable}_{0}$ of size $\mb{HashTable}_{1}.\mb{size} / 2$\\
            Relabel the hash tables with indices between $1$ to $k+1$\\
            \ForEach {$y \in \mb{HashTable}_{k+1}$} {
                \textsc{Insert($y$)}
            }
            Free $\mb{HashTable}_{k+1}$
        }
    }
\end{algorithm}

We now analyze the I/O cost $Q$ of the \ourht{}.  Here we assume that the size of the \ourht{} is larger than the \smallmem{} and $1-r<1/B$, so that one single lookup, insertion or deletion in a single level in the hash table on average requires no more than $c<2$ cache-line loads to find the location.

\myparagraph{Lookup.}  In a \ourht{}, a lookup requires $ck$ instead of $c$ \cm{}s ($c$ is the constant just defined) in the worst case (can quit earlier once the entry is found).  The cost increases by a factor of $k$ at most.

\myparagraph{Insert.}  There are two definitions of insertions: an insertion that the key is known to be not in the set/map, or an insertion that it is unknown whether the key is in this set/map.
Both cases are commonly-used. In this paper, we take the first definition
and analyze the cost of this type of insertions.
The second type of insertion can be viewed as a lookup first, then an insert if the lookup fails.

When inserting an element in a \ourht{}, we always try the smaller tables first.
Once all tables are full, we resize it.
More details can be found in Algorithm~\ref{algo:hash-table}.

The I/O cost $Q$ of an insertion comes in two parts: the cost of the initial insertion to the hash table, and the cost of this entry in future hash-table resizings.
The cost of the initial insertion is no more than $c+\wcost{}$, where $c$ is the number of cache-line reads to find the position to insert, plus $\wcost$, one cache-line write for the actual insertion.
The cost of resizing is more complicated to analyze.

We note that although a specific entry can be reinserted multiple times during different resizing processes, the overall number of element reinsertion is bounded, and thus we can a amortize the work.
A resizing occurs when an insertion comes in and the hash table contains exactly $r\cdot 2^p(2^k-1)$ elements for some positive integer $p$.
In this case, at most $r\cdot 2^p$ entries (the size of the smallest hash table), are reinserted during the resizing.
The total number of insertions from the last resizing is at least $r\cdot 2^{p-1}(2^k-1)$ (assuming $4l\le r$),  so the amortized I/O cost $Q$ of reinsertion for each insertion is upper bounded by $\displaystyle{(c+\wcost{})r\cdot 2^p \over r\cdot 2^{p-1}(2^k-1)}=(c+\wcost{})\cdot 2/(2^{k}-1)$.

In the asymmetric setting when $\wcost{} \gg 1$, the I/O cost of each insertion is approximately $\wcost{}\cdot(1+2/(2^{k}-1))$, indicating that compared to the classic implementation where $k=1$, in the worst-case the improvement when $k=2,3,4$ is about 44\%, 57\% and 62\% respectively.  The asymptotic improvement when $k\to +\infty$ is 67\%.

\myparagraph{Delete.}  A deletion in the \ourht{} is similar to an insertion except that a lookup for the location is required (details in Algorithm~\ref{algo:hash-table}).
The cost of the initial deletion is $ck+\wcost{}$.
A resizing of the hash table can occur after at least $l\cdot 2^p(2^k-1)$ deletions for some positive integer $p$, and the current hash table keeps $l\cdot 2^p(2^k-1)$ entries.
However, it is possible that all of these entries are in the last hash table so they are all reinserted.
We note that when reinserting the elements from the discarded array, we always try smaller arrays first.
This means that a reinserted entry, if not being deleted in the future, will not be reinserted again in the next $\min(k-1,\log_2{r/2l})$ shrinking resizings.
Namely, the amortized extra cost of a deletion in future resizings is about $\wcost{}/k$ if $l$ is set to be about $2^{-k}r$.
The overall I/O cost for a deletion is $Q=ck+\wcost{}(1+1/k)$.

\medskip
We have bounded of the I/O cost of each lookup, insertion or deletion, and the overall cost $Q$ can be estimated by summing the amount of each operation multiplied by the cost of this operation.
In practice, insertions and deletions can interleave.
For example, when a deletion comes after an insertion, the number of entries remains the same, which leads to no further cost for these two updates afterward.
The exact cost is also affected by the pattern of the sequence of the operations, and we will show by experiments.

\subsubsection{Experiments}

In the experiment we test the performance of our \ourht{}, including the numbers of \cm{s} and \wrt{s}, I/O costs, and wall-clock running time, on different patterns of queries.
In all experiments, we insert 1 million elements to an empty hash table, each of which is a 4-byte integer, into the hash table, and we vary the number of queries.
The simulated cache contains 10,000 cache-lines and uses the \emph{classic} policy, and for wall-clock running time we run the code on a PC with Intel i7-2600 CPU and 8GB RAM.
The occupancy rate is set to be $l=0.2$ and $r=0.8$.
We have tried other parameters ($r$ between 0.6 and 0.8 and $l=r/4$).
The results slightly vary, but all general conclusions in this section still hold.

\myparagraph{Non-deletion cases.}

\begin{figure*}[!h!t]
\small
\def\arraystretch{1.1}
  \centering
  $10^6$ insertions, $\alpha\times 10^6$ queries where $\alpha$ is from 0 to 8, cache size is 10,000 cache-lines.
    \begin{tabular}{p{.5cm}<{\centering}|p{.55cm}<{\raggedleft}r|p{.55cm}<{\raggedleft}r|p{.55cm}<{\raggedleft}r|p{.55cm}<{\raggedleft}r|p{.55cm}<{\raggedleft}r|
                                         p{.55cm}<{\raggedleft}r|p{.55cm}<{\raggedleft}r|p{.55cm}r}\toprule
     $\bm{\alpha}$& \multicolumn{2}{c|}{\textbf{0}} & \multicolumn{2}{c|}{\textbf{1/8}} & \multicolumn{2}{c|}{\textbf{1/4}} & \multicolumn{2}{c|}{\textbf{1/2}} & \multicolumn{2}{c|}{\textbf{1}} & \multicolumn{2}{c|}{\textbf{2}} & \multicolumn{2}{c|}{\textbf{4}} & \multicolumn{2}{c}{\textbf{8}} \\
     \hline
          & \multicolumn{1}{c}{\textbf{RT}} & \multicolumn{1}{c|}{\textbf{WT}} & \multicolumn{1}{c}{\textbf{~RT}} & \multicolumn{1}{c|}{\textbf{WT}} & \multicolumn{1}{c}{\textbf{~RT}} & \multicolumn{1}{c|}{\textbf{WT}} & \multicolumn{1}{c}{\textbf{~RT}} & \multicolumn{1}{c|}{\textbf{WT}} & \multicolumn{1}{c}{\textbf{~RT}} & \multicolumn{1}{c|}{\textbf{WT}} & \multicolumn{1}{c}{\textbf{~RT}} & \multicolumn{1}{c|}{\textbf{WT}} & \multicolumn{1}{c}{~~RT} & \multicolumn{1}{c|}{\textbf{WT}} & \multicolumn{1}{c}{\textbf{~RT}} & \multicolumn{1}{c}{\textbf{WT}} \\\hline
    \textbf{k=1}     & 1.35  & 1.17  & 1.44  & 1.18  & 1.52  & 1.19  & 1.69  & 1.21  & 2.02  & 1.24  & 2.68  & 1.27  & 4.00  & 1.31  & 6.64  & 1.34 \\
    \textbf{k=2}     & 0.85  & 0.79  & 1.06  & 0.84  & 1.23  & 0.87  & 1.54  & 0.91  & 2.09  & 0.96  & 3.11  & 1.00  & 5.07  & 1.03  & 8.94  & 1.05 \\
    \textbf{k=3}     & 0.76  & 0.72  & 1.08  & 0.80  & 1.32  & 0.85  & 1.73  & 0.90  & 2.44  & 0.95  & 3.76  & 0.99  & 6.31  & 1.02  & 11.32 & 1.05 \\
    \textbf{k=4}     & 0.70  & 0.67  & 1.11  & 0.78  & 1.40  & 0.82  & 1.89  & 0.88  & 2.74  & 0.93  & 4.30  & 0.97  & 7.33  & 1.00  & 13.31 & 1.03 \\
    \bottomrule
    \end{tabular}%
  \captionof{table}{Numbers of \rwt{s} of \ourht{s} with different query/insert ratios.  Numbers of \rwt{s} are divided by 1M.}
  \label{tbl:hashtable-raw}%
  \medskip
  The I/O costs of \ourht{s} with the same settings in Table~\ref{tbl:hashtable-raw}.
    \begin{tabular}{p{.5cm}<{\centering}|r@{ }@{ }@{ }r@{ }@{ }@{ }r@{ }@{ }@{ }r@{ }@{ }@{ }r@{ }@{ }@{ }r@{ }@{ }@{ }r@{ }@{ }@{ }r|r@{ }@{ }@{ }r@{ }@{ }@{ }r@{ }@{ }@{ }r@{ }@{ }@{ }r@{ }@{ }@{ }r@{ }@{ }@{ }r@{ }@{ }@{ }r}\toprule
     & \multicolumn{8}{c|}{$\bm{\wcost{}=10}$}                                      & \multicolumn{8}{c}{$\bm{\wcost{}=100}$} \\
\hline
          $\bm{\alpha}$& \multicolumn{1}{c}{\textbf{0}} & \multicolumn{1}{c}{\textbf{1/8}} & \multicolumn{1}{c}{\textbf{1/4}} & \multicolumn{1}{c}{\textbf{1/2}} & \multicolumn{1}{c}{\textbf{1}} & \multicolumn{1}{c}{\textbf{2}} & \multicolumn{1}{c}{\textbf{4}} & \multicolumn{1}{c|}{\textbf{8}} & \multicolumn{1}{c}{\textbf{0}} & \multicolumn{1}{c}{\textbf{1/8}} & \multicolumn{1}{c}{\textbf{1/4}} & \multicolumn{1}{c}{\textbf{1/2}} & \multicolumn{1}{c}{\textbf{1}} & \multicolumn{1}{c}{\textbf{2}} & \multicolumn{1}{c}{\textbf{4}} & \multicolumn{1}{c}{\textbf{8}} \\
\hline
    \textbf{k=1}     & 13.0 & 13.2 & 13.4 & 13.8 & 14.4 & 15.4 & 17.1 & 20.0 & 117.9 & 119.3 & 120.5 & 122.7 & 125.8 & 129.9 & 134.8 & 140.5 \\
    \textbf{k=2}     & \textcolor[rgb]{ 0,  0,  1}{8.8} & \textcolor[rgb]{ 0,  0,  1}{9.5} & \textcolor[rgb]{ 0,  0,  1}{10.0} & \textcolor[rgb]{ 0,  0,  1}{10.7} & \textcolor[rgb]{ 1,  0,  0}{\underline{11.7}} & \textcolor[rgb]{ 1,  0,  0}{\underline{13.1}} & \textcolor[rgb]{ 1,  0,  0}{\underline{15.4}} & \textcolor[rgb]{ 1,  0,  0}{\underline{19.5}} & \textcolor[rgb]{ 0,  0,  1}{79.9} & \textcolor[rgb]{ 0,  0,  1}{85.1} & \textcolor[rgb]{ 0,  0,  1}{88.4} & \textcolor[rgb]{ 0,  0,  1}{92.8} & \textcolor[rgb]{ 0,  0,  1}{97.7} & \textcolor[rgb]{ 0,  0,  1}{102.9} & \textcolor[rgb]{ 0,  0,  1}{108.2} & \textcolor[rgb]{ 1,  0,  0}{\underline{114.4}} \\
    \textbf{k=3}     & \textcolor[rgb]{ 0,  0,  1}{8.0} & \textcolor[rgb]{ 0,  0,  1}{9.1} & \textcolor[rgb]{ 0,  0,  1}{9.8} & \textcolor[rgb]{ 1,  0,  0}{\underline{10.7}} & \textcolor[rgb]{ 0,  0,  1}{11.9} & \textcolor[rgb]{ 0,  0,  1}{13.7} & \textcolor[rgb]{ 0,  0,  1}{16.5} & 21.8 & \textcolor[rgb]{ 0,  0,  1}{73.1} & \textcolor[rgb]{ 0,  0,  1}{81.4} & \textcolor[rgb]{ 0,  0,  1}{85.8} & \textcolor[rgb]{ 0,  0,  1}{91.3} & \textcolor[rgb]{ 0,  0,  1}{97.0} & \textcolor[rgb]{ 0,  0,  1}{102.7} & \textcolor[rgb]{ 0,  0,  1}{108.7} & \textcolor[rgb]{ 0,  0,  1}{116.3} \\
    \textbf{k=4}     & \textcolor[rgb]{ 1,  0,  0}{\underline{7.4}} & \textcolor[rgb]{ 1,  0,  0}{\underline{8.9}} & \textcolor[rgb]{ 1,  0,  0}{\underline{9.6}} & \textcolor[rgb]{ 0,  0,  1}{10.7} & \textcolor[rgb]{ 0,  0,  1}{12.0} & \textcolor[rgb]{ 0,  0,  1}{14.0} & 17.4 & 23.6 & \textcolor[rgb]{ 1,  0,  0}{\underline{67.9}} & \textcolor[rgb]{ 1,  0,  0}{\underline{78.6}} & \textcolor[rgb]{ 1,  0,  0}{\underline{83.8}} & \textcolor[rgb]{ 1,  0,  0}{\underline{89.6}} & \textcolor[rgb]{ 1,  0,  0}{\underline{95.5}} & \textcolor[rgb]{ 1,  0,  0}{\underline{101.3}} & \textcolor[rgb]{ 1,  0,  0}{\underline{107.7}} & \textcolor[rgb]{ 0,  0,  1}{116.1} \\
    \bottomrule
    \end{tabular}%
  \caption{The I/O costs of \ourht{s} with different query/insert ratios.  The write-read ratio $\wcost{}$ are selected to be typical projected values 10 (latency, bandwidth) and 100 (energy).  Results are based on the numbers in Table~\ref{tbl:hashtable-raw}.  Numbers in red with underlines indicate the best choice of $k$ that minimizes the I/O cost in this setting, and numbers in blue indicate better I/O costs comparing to the classic hash table implementation (i.e.\ $k=1$).}
  \label{tab:hashtable-weighted}%
\begin{center}
    \begin{minipage}[t]{0.45\textwidth}
        \includegraphics[width=\textwidth]{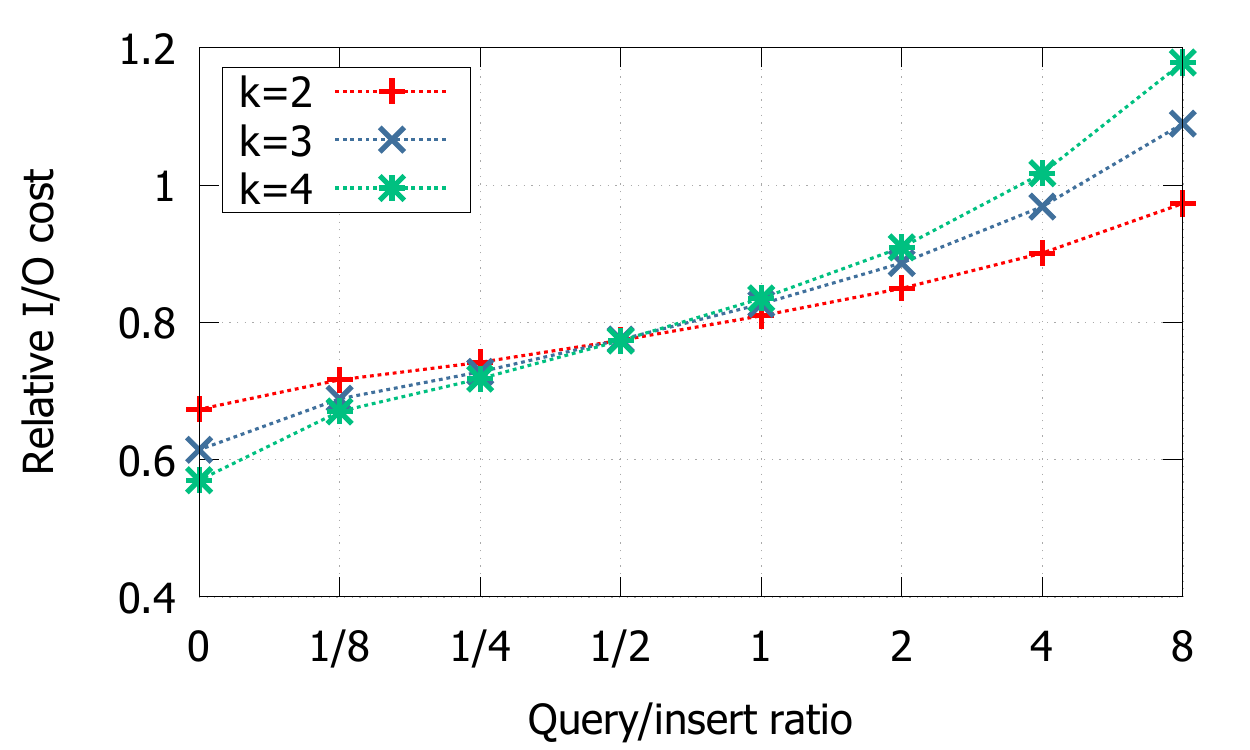}\\ \centering{\small (a) $\wcost=10$}
    \end{minipage}~~~
    \begin{minipage}[t]{0.45\textwidth}
        \includegraphics[width=\textwidth]{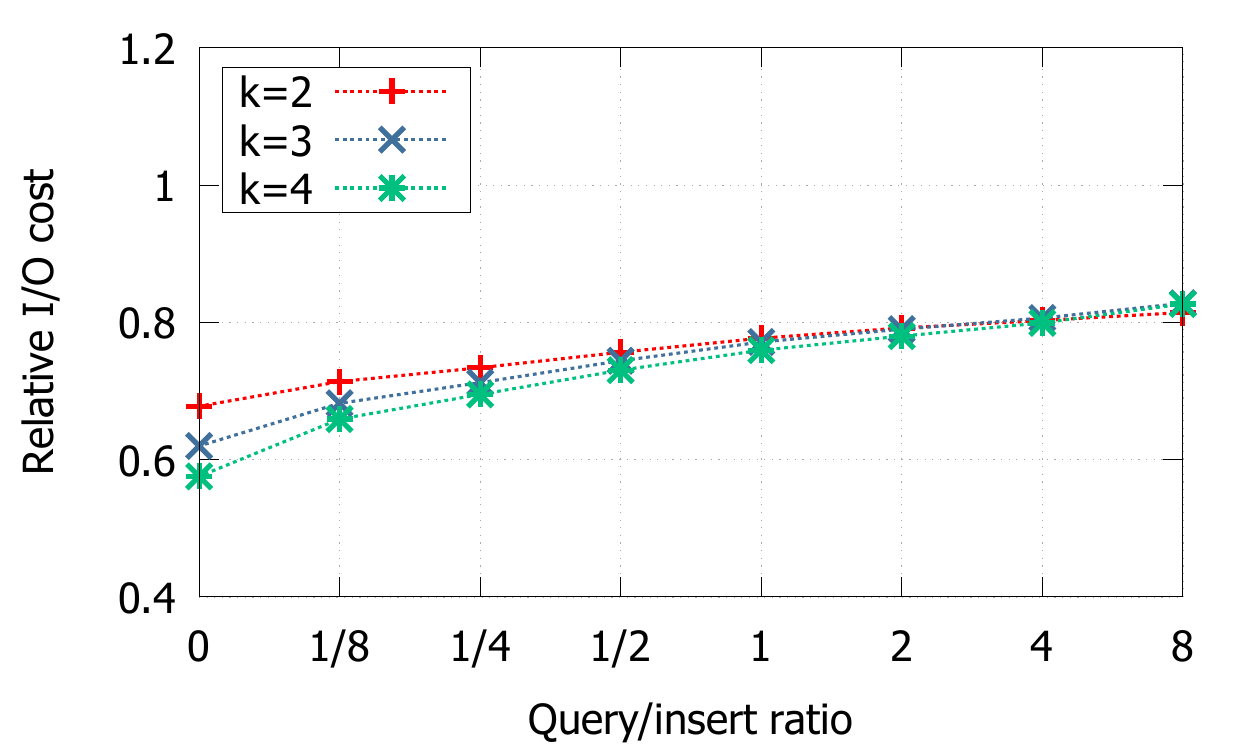}\\ \centering{\small (b) $\wcost=100$}
    \end{minipage}
\vspace{.5em}
    \captionof{figure}{Relative I/O cost of \ourht{} with different $k$.  The I/O cost is divided by the $k=1$ case, so every data point below 1 indicates an improvement in such case. Numbers are from Table~\ref{tab:hashtable-weighted}.}\label{fig:hashtable}
\end{center}
\end{figure*}

Many applications, like webpage caching or the breadth-first searches, only insert but never delete elements in a hash table.
Our experiment starts with this simpler case. 
We first show the relationship between $k$ (the number of hash tables) and the numbers of \cm{s} and \wrt{s} for a variety of insertion/query ratios, and the results are shown in Table~\ref{tbl:hashtable-raw}.
We fix the number of insertions to be one million, and query $\alpha$ times after each insertion.
We vary $\alpha$ from 0, 1/8, to 8 ($\alpha<1$ indicates one query per $1/\alpha$ insertions).
About 50\% query keys are in the hash table (this ratio affects the I/O cost since a successful query can terminate earlier).
The number of levels $k$ varies from 1 to 4.
In Table~\ref{tab:hashtable-weighted}, we show the overall I/O costs, which are the weighted sums assuming two typical values of the write-read ratio $\wcost$, 10 and 100.

We first look at the number of \wrt{s}.
When there is no query (i.e., the first column, just inserting 1 million entries), the numbers of writes are consistent with our analysis for insertions in Section \ref{sec:hashtable}.
The only exception here is that cache can hold a constant fraction of the elements, which batches the writes and reduces the number of memory transfers.
However, the relative trend in each column remains unchanged.
Namely, the number of writes always decreases as the increase of $k$ regardless of the ratio between queries and updates.
The number of writes is reduced by $33\%$, $40\%$ and $43\%$ when $k=2,3,4$ respectively. 
Such improvement also shows up in the overall I/O cost in Table~\ref{tab:hashtable-weighted}.

We note that more queries cause more reads, and larger $k$ also leads to more reads.
Since these reads flush the cache-lines, the numbers of writes in these cases also marginally increase.
The optimal choice of $k$ is decided by the update/query distribution as well as the write-read ratio $\wcost$.
In general, more queries lead to worse performance with larger $k$, and larger $\wcost$ prefers larger $k$.
In Table~\ref{tab:hashtable-weighted}, we underline the numbers indicating the best choice of $k$ in that specific setting.
The experiment results indicate that picking $k$ to be 2 or 3 is always a good choice when $\wcost=10$, and 3 or 4 when $\wcost=100$.

\myparagraph{Wall-clock running time.}

We also measure the actual running time of the previously stated operations on a real machine.
In the current platform with symmetric memory-access costs, the write-read ratio is close to 1.
Here we show that, the weighted sums of the I/O measures in Table~\ref{tbl:hashtable-raw} almost match the actual running times shown in Table~\ref{tbl:wallclock}, when plugging $\wcost=1$.
Therefore, it is reasonable to believe that, the I/O cost shown in Table~\ref{tab:hashtable-weighted} can reasonably well project the performance for the future memory, when the bandwidths for reads and writes become asymmetric. (The argument for energy consumptions holds independently with this running time and other architectural issues.)
Meanwhile, it is interesting to point out that, when insertions are more than queries, using \ourht{} with $k=2$ is actually faster even in the current platform.

Our implementation has a much lower I/O cost compared to separate chaining.  We run the same experiment using the STL unordered set (with the same hash function and other setups), and our hash table is at least 3-4 times faster in all cases.

\begin{table}[!ht]
\small
\def\arraystretch{1.1}
  \centering
  Wall-clock running time (milliseconds), $10^6$ insertions, $\alpha\times 10^6$ queries where $\alpha$ is from 0 to 8.
    \begin{tabular}{p{.5cm}<{\centering}|rrrrrrrr}
    \toprule
    $\bm{k}$   & \multicolumn{1}{c}{\textbf{0}} & \multicolumn{1}{c}{\textbf{1/8}} & \multicolumn{1}{c}{\textbf{1/4}} & \multicolumn{1}{c}{\textbf{1/2}} & \multicolumn{1}{c}{\textbf{1}} & \multicolumn{1}{c}{\textbf{2}} & \multicolumn{1}{c}{\textbf{4}} & \multicolumn{1}{c}{\textbf{8}} \\\hline
    \textbf{1}     & 61    & 65    & 69    & 77    & 95    & 129   & 191   & 321 \\
    \textbf{2}     & \textcolor[rgb]{ 0,  0,  1}{36} & \textcolor[rgb]{ 0,  0,  1}{45} & \textcolor[rgb]{ 0,  0,  1}{53} & \textcolor[rgb]{ 0,  0,  1}{69}    & 96    & 163   & 277   & 512 \\
    \textbf{3}     & \textcolor[rgb]{ 0,  0,  1}{31} & \textcolor[rgb]{ 0,  0,  1}{45} & \textcolor[rgb]{ 0,  0,  1}{57}    & 80    & 124   & 213   & 392   & 739 \\
    \textbf{4}     & \textcolor[rgb]{ 0,  0,  1}{30} & \textcolor[rgb]{ 0,  0,  1}{51} & \textcolor[rgb]{ 0,  0,  1}{66}    & 95    & 156   & 278   & 519   & 999 \\
    \bottomrule
    \end{tabular}%
  \caption{Wall-clock running time (milliseconds) of \ourht{} with different combinations of $k$ and query/insert ratios.   Numbers in blue indicate a better performance comparing to the classic implementation (i.e.\ $k=1$).}
  \label{tbl:wallclock}%
\medskip
  1 million insertions then 1 million deletions, $\alpha$ times 2 million queries where $\alpha$ is from 0 to 4, 10,000 cache-lines.
    \begin{tabular}{p{.4cm}<{\centering}|r@{ }@{ }@{ }r|r@{ }@{ }@{ }r|r@{ }@{ }@{ }r|r@{ }@{ }@{ }r}\toprule
    \multirow{2}[0]{*}{$\bm{k}$} & \multicolumn{2}{c|}{\textbf{0}} & \multicolumn{2}{c|}{\textbf{1/4}} & \multicolumn{2}{c|}{\textbf{1}} & \multicolumn{2}{c}{\textbf{4}} \\\cline{2-9}
          & \multicolumn{1}{c}{\textbf{RT}} & \multicolumn{1}{c|}{\textbf{WT}} & \multicolumn{1}{c}{\textbf{RT}} & \multicolumn{1}{c|}{\textbf{WT}} & \multicolumn{1}{c}{\textbf{RT}} & \multicolumn{1}{c|}{\textbf{WT}} & \multicolumn{1}{c}{\textbf{RT}} & \multicolumn{1}{c}{\textbf{WT}} \\\hline
    \textbf{1}     & 2.55  & 2.32  & 2.95  & 2.36  & 4.14  & 2.44  & 8.91  & 2.52 \\
    \textbf{2}     & 2.28  & 1.75  & 3.15  & 1.88  & 5.45  & 2.04  & 14.14 & 2.18 \\
    \textbf{3}     & 2.47  & 1.59  & 3.79  & 1.80  & 7.11  & 1.98  & 19.61 & 2.12 \\
    \textbf{4}     & 2.72  & 1.50  & 4.44  & 1.75  & 8.68  & 1.95  & 24.67 & 2.08 \\
    \bottomrule
    \end{tabular}%
  \caption{Numbers of \rwt{s} of \ourht{s} with different query/(insert+delete) ratios.  Numbers of \rwt{s} are devided by 1M.}
  \label{tbl:hashtable-delete-raw}%
\medskip
  The I/O costs of \ourht{s} with the same settings in Table~\ref{tbl:hashtable-delete-raw}.
    \begin{tabular}{p{.4cm}<{\centering}|rrrr|rrrr}\toprule
    \multicolumn{1}{c|}{\multirow{2}[0]{*}{$\bm{k}$}} & \multicolumn{4}{c|}{$\bm{\wcost{}=10}$}     & \multicolumn{4}{c}{$\bm{\wcost{}=100}$} \\\cline{2-9}
          & \multicolumn{1}{c}{\textbf{0}} & \multicolumn{1}{c}{\textbf{1/4}} & \multicolumn{1}{c}{\textbf{1}} & \multicolumn{1}{c|}{\textbf{4}} & \multicolumn{1}{c}{\textbf{0}} & \multicolumn{1}{c}{\textbf{1/4}} & \multicolumn{1}{c}{\textbf{1}} & \multicolumn{1}{c}{\textbf{4}} \\\hline
    \textbf{1}     & 25.8  & 26.6  & 28.5  & \textcolor[rgb]{ 1,  0,  0}{\underline{34.2}} & 235.0 & 239.4 & 247.6 & 261.4 \\
    \textbf{2}     & \textcolor[rgb]{ 0,  0,  1}{19.7} & \textcolor[rgb]{ 0,  0,  1}{22.0} & \textcolor[rgb]{ 1,  0,  0}{\underline{25.9}} & \textcolor[rgb]{ 0,  0,  1}{35.9} & \textcolor[rgb]{ 0,  0,  1}{176.8} & \textcolor[rgb]{ 0,  0,  1}{191.6} & \textcolor[rgb]{ 0,  0,  1}{209.6} & \textcolor[rgb]{ 0,  0,  1}{232.0} \\
    \textbf{3}     & \textcolor[rgb]{ 0,  0,  1}{18.4} & \textcolor[rgb]{ 1,  0,  0}{\underline{21.8}} & \textcolor[rgb]{ 0,  0,  1}{26.9} & \textcolor[rgb]{ 0,  0,  1}{40.8} & \textcolor[rgb]{ 0,  0,  1}{161.4} & \textcolor[rgb]{ 0,  0,  1}{183.6} & \textcolor[rgb]{ 0,  0,  1}{205.1} & \textcolor[rgb]{ 1,  0,  0}{\underline{231.4}} \\
    \textbf{4}     & \textcolor[rgb]{ 1,  0,  0}{\underline{17.7}} & \textcolor[rgb]{ 0,  0,  1}{22.0} & \textcolor[rgb]{ 0,  0,  1}{28.2} & \textcolor[rgb]{ 0,  0,  1}{45.5} & \textcolor[rgb]{ 1,  0,  0}{\underline{152.3}} & \textcolor[rgb]{ 1,  0,  0}{\underline{179.6}} & \textcolor[rgb]{ 1,  0,  0}{\underline{203.5}} & \textcolor[rgb]{ 0,  0,  1}{233.1} \\
    \bottomrule
    \end{tabular}%
  \caption{The I/O costs of \ourht{s} with different query vs.\ insert/delete ratios.  The write-read ratio $\wcost{}$ are 10 and 100.  Results are based on the numbers in Table~\ref{tbl:hashtable-delete-raw}.  Numbers in red with underlines indicate the best choice of $k$ that minimizes the I/O cost, and numbers in blue indicate better I/O costs comparing to the classic hash table implementation (i.e.\ $k=1$).}
\bigskip
  \label{tab:hashtable-delete-weighted}%
\end{table}%

\myparagraph{Insertion and deletion cases.}

We also tested the performance of \ourht{} on deletions.
We first insert 1 million elements and then remove them all.
After each insertion or deletion, we query $\alpha$ times. The other settings are the same as the non-deletion case.
The results on the numbers of \cm{} and \wrt{} are shown in Table~\ref{tbl:hashtable-delete-raw}, and the overall I/O cost with write-read ratio $\wcost$ to be 10 and 100 are shown in Table~\ref{tab:hashtable-delete-weighted}.

From the results, we get almost the same trend as the non-deletion cases.
Compared to the classic implementation (i.e., $k=1$), the overall number of \wrt{} is reduced by 25\%, 32\%, and 36\% when no queries are involved, and the improvement is slightly decreased when more queries come in, since more read accesses flush out the cache-lines.
We note that the number of \cm{s} required by a deletion is more than that for an insertion, since in each deletion we need to locate the element in the hash table, which requires to look up in most $k$ hash table levels.
Hence, compared to the non-deletion cases, slightly smaller values for $k$ are more preferable.
The best choice of $k$ in each case is underlined and shown in Table~\ref{tab:hashtable-delete-weighted}.
For smaller $\wcost{}=10$ (indicating bandwidth and latency), the best choice of $k$ varies based on different query/update ratios, but $k=2$ is always an acceptable choice.
When considering the energy consumption ($\wcost\approx 100$), a larger $k$, like 3 or 4, is more desirable in all cases.

\subsubsection{Conclusions}

We proposed a new data structure, the \ourht{}, to implement unordered set and map, that has the same space utilization compared to the classic open-addressing hash tables.
The key idea in the \ourht{} is to keep multiple instead of one level of hash tables.
As a result, the algorithm uses fewer writes during resizings, at the cost of more reads in other operations.

The best choice of $k$ is decided by the ratio of updates and queries.
Our experiment shows that $k=2$ always leads to a lower or similar I/O cost when the query/insert ratio is no more than 8, compared to the classic $k=1$ setting.
For the ratio of write/read cost is larger (like $100$), larger values of $k$, like 3 or 4, are even more preferable than the $k=2$ case.

\hide{
\myparagraph{Breadth-first search as an application.}
In the full version of this paper, we discussed our implementations on breadth-first searches (BFS) on graph traversing or searching.
Our algorithms compute the single-source shortest paths (SSSP) or s-t shortest-paths on unweighted graphs, which can further apply to graph radii estimation, eccentricity estimation and betweenness centrality, and as a basic building block for other graph algorithms like graph connectivity, reachability, bridges, biconnected components, and strongly connected components.

There are two technical contributions in our new algorithms: one is the new algorithm design that always constrains the size of the temporal memory used in the search to be the (order of the) maximum size of the any frontier (i.e.\ the set of vertices with the same distance to the source node); the other is the new \ourht{} that acts as the underlying data structure to run BFS, which minimizes the number of writes to the main memory.

With the new BFS algorithm, we show a significant improvement on eight real-world graphs (details of these graphs are provided in Section~\ref{sec:dijk-exp}) with various cache sizes, compared to the classic queue-based implementations.
When $\wcost=10$, we show an up to 8-fold improvement on SSSP, and up to 43-fold improvement on s-t shortest-paths.
For $\wcost=100$, the improvement is more stable, which is 69 on SSSP and 71 on s-t shortest-paths.
Meanwhile, the new algorithm requires less I/O cost in almost all combinations of input instances and cache parameters.
}

\hide{
In the experiment we test the performance of our \ourht{}, including the numbers of \cm{s} and \wrt{s}, I/O costs, and wall-clock running time, on different patterns of queries.
In all experiments, we insert 1 million elements, each of which is a 4-byte integer, into the hash table, and we vary the number of queries.
The simulated cache contains 10,000 cache-lines, and for wall-clock running time we run the code on a PC with Intel i7-2600 CPU and 8GB RAM.
The occupancy rate is set to be $l=0.2$ and $r=0.8$.
We tried other parameters ($r$ between 0.6 and 0.8 and $l=r/4$), and the results are slightly changed but all conclusions drawn in this section remains unchanged.

\myparagraph{Non-deletion case.}
A number of applications, like webpage caching or the breadth-first searches, only insert but never delete elements in a hash table.
Since this is a simpler case, our experiment starts here without considering deletions.
We first show the relationship between $k$ (the number of hash tables) and the numbers of \cm{s} and \wrt{s} for a variety of insertion / query ratio, and the result is shown in Table~\ref{tbl:hashtable-raw}.
We fix the number of insertions to be one million, and query $\alpha$ times after each insertion. We vary $\alpha$ from 0, 1/8 million, ..., to 8 million.
The ratio of hit is about 50\% (this ratio affects the I/O cost since a successful query can terminate earlier).
The number of hash tables $k$ varies from 1 to 4.
In Table~\ref{tab:hashtable-weighted}, we show the overall I/O costs, which are the weighted sums assuming two typical values of the write-read ratio $\wcost$, 10 and 100.

We first look at the number of \wrt{s}.
When there is no query (i.e., the first column, just inserting 1 million entries), the numbers of writes are consistent with our analysis for insertions in Section \ref{sec:hashtable}.
The only exception here is that cache can hold a constant fraction of the elements, which batches the writes and reduces the number of transfers.
However, the relative trend across different rows (i.e.\ different $k$) remains unchanged.
Namely, the number of writes always decreases as the increase of $k$ regardless of the ratio between queries and updates.
The percentage of writes is reduced by $33\%$, $40\%$ and $43\%$ when $k=2,3,4$ respectively.
This reduction also shows up in the overall I/O cost in Table~\ref{tab:hashtable-weighted}.

More queries clearly cause more reads, and larger $k$ leads to more reads (since these reads flush the cache-lines, the overall number of writes also increases by the increase of the queries).
The optimal I/O cost is affected by $k$, input distribution, and the write-read ratio $\wcost$.
More queries lead to worse performance with larger $k$, and larger $\wcost$ prefers larger $k$.
In Table~\ref{tab:hashtable-weighted} we underline the numbers indicating the best choice of $k$ in that setting.
Picking $k$ to be 2 or 3 is always a good choice when $\wcost=10$, and 3 or 4 when $\wcost=100$.

\myparagraph{Wall-clock running time.}
We also measure the actual running time of the previously stated operations on a PC.
In our current platform and hardware, the write-read ratio is close to 1.
Here we show that, the numbers in Table~\ref{tbl:hashtable-raw} almost match the actual running time shown in Table~\ref{tbl:wallclock}, when plugging $\wcost=1$.
Therefore, it is reasonable to believe that, the I/O cost shown in Table~\ref{tab:hashtable-weighted} can reasonably well project the performance for the future memory, when the bandwidth changes asymmetrically (the argument for energy consumptions holds independently with this running time and other architectural issues).
Meanwhile, in the current platform, if the number of insertion is more than the number of queries, using \ourht{} with $k=2$ improves the performance.

As we stated at the beginning of this section, in this experiment setting, our implementation has a much lower I/O cost comparing to separate chaining.  We run the same experiment using the STL unordered set (with the same hash function and other setups), and our hash table is at least 3-4 times faster in all cases.

\myparagraph{Insertion and deletion case.}
We also tested the performance of \ourht{} on deletions.
We first insert 1 million elements and then remove them all.
After each insertion or deletion, we query $\alpha$ times. The other settings are the same as the non-deletion case.
The results on the numbers of \cm{} and \wrt{} are shown in Table~\ref{tbl:hashtable-delete-raw}, and the overall I/O cost with write-read ratio $\wcost$ to be 10 and 100 are shown in Table~\ref{tab:hashtable-delete-weighted}.

From the results, we get almost the same trend as the non-deletion case.
Comparing with the classic implementation (i.e., $k=1$), the overall number of \wrt{} is reduced by 25\%, 32\% and 36\% when no queries involved, and the improvement is slightly decreased when more queries come in, since more read accesses flush out the cache-lines.
Notice that, the number of \cm{} required by a deletion is more than that for an insertion, since in each deletion we need to allocate the element in the hash table, which requires to look up at most $k$ cache-lines.
Hence comparing to the non-deletion case, slightly smaller values for $k$ are more preferable.
The best choice of $k$ in each case is underlined and shown in Table~\ref{tab:hashtable-delete-weighted}.
For smaller $\wcost{}=10$ (indicate bandwidth and latency) the best $k$ varies based on different query/update ratios, but $k=2$ is always an acceptable choice.
When considering the energy consumption ($\wcost\approx 100$), a larger $k$, like 3 or 4, is more desirable in all cases.
}
\hide{
\myparagraph{Breadth-first search as an application.}
In the full version of this paper we discussed our implementations on breadth-first searches (BFS) on graph traversing or searching.
Our algorithms computes the single-source shortest paths (SSSP) or s-t shortest-paths on unweighted graphs, which can further apply to graph radii estimation, eccentricity estimation and betweenness centrality, and as a basic building block for other graph algorithms like graph connectivity, reachability, bridges, biconnected components, and strongly connected components.

There are two technical contributions in our new algorithms: one is the new algorithm design that always constrains the size of the temporal memory used in the search to be the (order of the) maximum size of the any frontier (i.e.\ the set of vertices with the same distance to the source node); the other is the new \ourht{} that acts as the underlying data structure to run BFS, which minimizes the number of writes to the main memory.

With the new BFS algorithm, we show a significant improvement on eight real-world graphs (details of these graphs are provided in Section~\ref{sec:dijk-exp}) with various cache sizes, compared to the classic queue-based implementations.
When $\wcost=10$, we show an up to 8-fold improvement on SSSP, and up to 43-fold improvement on s-t shortest-paths.
For $\wcost=100$, the improvement is more stable, which is 69 on SSSP and 71 on s-t shortest-paths.
Meanwhile, the new algorithm requires less I/O cost in almost all combinations of input instances and cache parameters.
}

\subsection{Ordered Sets and Maps}
\label{sec:orderedset}

The implementations of ordered sets and maps are based on some form
of balanced tree (or tree-like) data structure and, at minimum, support \textbf{lookup}, \textbf{insertion}, and \textbf{deletion} in logarithmic time.
Some set-set functions such as \textbf{union}, \textbf{intersection}, and \textbf{difference} are also required in many scenarios.

One commonly-used data structure to maintain ordered sets and maps is the self-balancing binary search tree (BST).  Most languages and libraries use either AVL tree~\cite{adelsonvelskii1963algorithm} or red-black tree~\cite{bayer1972symmetric,guibas1978Dichromatic} to implement them, while some other implementations of weight-balanced trees~\cite{nievergelt1973binary} and treaps~\cite{Aragon89,seidel1996randomized} also exist.  Here we first analyze the I/O cost on some existing solutions.

\subsubsection{I/O cost on BSTs}
\label{sec:bst-single}

For simplicity, here we assume that the \smallmem{} size is $M=O(1)$.
Locating the key for a lookup, insertion or deletion requires to load and compare to $\Theta(\log n)$ tree nodes on the balanced binary search trees (BSTs), The I/O cost is $\Theta(\log n)$, which is also the lower bound of such operations.

For an insertion or deletion, we also need to modify the tree accordingly.
In the asymmetric setting, weight-balanced trees~\cite{nievergelt1973binary} are not a good option since we have to update the subtree sizes all the way to the root.
This update leads to $\Theta(\log n)$ writes to the \largemem{} per update.
For the other types of BSTs, we show their I/O costs on insertions and deletions individually.

\myparagraph{Red-black trees.}  Red-black trees~\cite{bayer1972symmetric,guibas1978Dichromatic} have the simplest update rules among these balanced BSTs.  With the classic rebalancing rules and careful implementation, it requires only $O(1)$ amortized time per update (insertion or deletion) after locating the key~\cite{Tarjan83}. 
Also, red-black trees require no extra cost to update balancing information except for the tree nodes involved in rotations (unlike the case in AVL trees).
As a result, red-black trees have an optimal amortized I/O cost $Q=\Theta(\log n)$ per a lookup and $Q=\Theta(\wcost{}+\log n)$ per insertion or deletion on the \ourmodel{}.

\myparagraph{AVL trees.}  An insertion in AVL trees requires at most two rotations (a double rotation)~\cite{adelsonvelskii1963algorithm}.  Unlike the red-black trees however, we need to track and update the balance factors along the path from the root to the modified tree node.
We now bound the number of updated balance factors to be a constant.
If we store the height of the subtree in each tree node, the difference of the two subtree heights can be checked in constant time.
Since the number of subtrees of height $d$ in an AVL tree is no more than $n/c^{\lfloor d/2\rfloor}$ for a tree with $n$ nodes and some constant $c>1$~\cite{blelloch2016just}, the number of increments of the counts for $n$ nodes is $\sum_{d\ge 1}{d\cdot (n/c^{\lfloor d/2\rfloor})}=O(n)$.
On average, an insertion needs $O(n)/n=O(1)$ writes.

The deletions of AVL trees is more complicated and $\Theta(\log n)$ rotations can be applied on every deletion.
Amani et al.~\cite{amani2016amortized} recently showed that there exists such a sequence of $3n$ intermixed insertions and deletions on an initially empty AVL tree that takes $\Theta(n\log n)$ rotations.
This instance indicates that the classic implementation of AVL trees has an I/O cost $Q=\Theta(\wcost{}\log n)$ per deletion in the worst case.

\myparagraph{Treaps.}  A treap, also called as a randomized search tree~\cite{Aragon89,seidel1996randomized}, is a Cartesian tree in which each key is given a randomly chosen numeric priority, and the inorder traversal order of the nodes is the same as the sorted order of the keys.  The priority for any non-leaf node must be greater than or equal to the priority of its children.

When inserting an element into a treap with $n-1$ elements or removing an element from $n$ elements, The updated element only compares to the elements that each has a higher priority than the other elements between this element and the updated element.
The number of such comparisons is $\sum_{j\in [n]\backslash \{i\}}{1/\lvert i-j\rvert}=O(\log n)$ in expectation\footnote{Also with high probability.}, where $i$ is the position of the updated element in the total order of $n$ elements~\cite{seidel1996randomized}.

The number of rotations can be computed similarly.
For an insertion, a rotation happens once the inserted element has a higher priority than all elements in the entire subtree.
Again if we assume the inserted element ranked $i$-th in the total order, the probability that it rotates up for the $j$-th ranked tree node is $1/\lvert i-j\rvert^2$ (i.e., the $j$-th element has higher priority than all elements between, and lower than the priority of the inserted node).
The overall expected number of rotations per insertion is $\sum_{j\in [n]\backslash \{i\}}{1/\lvert i-j\rvert^2}=O(1)$ in expectation.
We can show the constant writes per deletion accordingly.

We note that unlike an AVL tree or a red-black tree, a treap does not require updates to the balancing criteria, which means that we never need to modify the information in each tree node after it is inserted.
As a result, an insertion, deletion or query on a treap requires $O(\log n)$ reads \whp{} and an insertion or deletion requires $O(1)$ writes in expectation.

\subsubsection{Join-based implementation}
\label{sec:join}

Blelloch et al.~\cite{blelloch2016just,sun2018pam} recently proposed a framework that supports fast bulk operations.
This framework supports efficient single or multiple insertions or deletions on AVL trees, red-black trees, treaps and weight-balanced trees, as well as union, intersection and set difference on two BSTs (of the same kind).
This framework is very concise: each function can be implemented within a dozen lines of code and are independent with the specific balancing criteria in different types of BSTs.
The functions rely on only one helper operation \JOIN{}~\cite{adams1992implementing,adams1993functional,ST85,Tarjan83,blelloch2016just}, which deals with the tree balancing and can be treated as a black box.
Under this framework, the implementation of these functions is highly efficient and parallelized.

\myparagraph{\JOIN{$(T_L,k,T_R)$}}: The \JOIN{} function takes two balanced BSTs (of the same balancing schemes) $T_L$ and $T_R$ and a key $k$, and returns a new balanced BST for which the in-order values are a concatenation of the in-order values of $T_L$, then $k$, and then the in-order values of $T_R$.
In order to keep the ordering in the tree nodes, the \JOIN{} function assumes $k$ to be greater than all keys in $T_L$ and smaller than all keys in $T_R$.
\JOIN{} handles all issues related to rebalancing, and is the only function that knows about and maintains the balance invariants.
The \JOIN{} algorithms for each of the balancing schemes are given in~\cite{blelloch2016just}.
Meanwhile, \JOIN{} is the only function that \emph{writes} to the memory.
More specifically, all operations that change the attributes of a tree node (e.g., linking to new children, height-maintaining for AVL or red-black trees, color-changing in red-black trees, etc.) are restricted in \JOIN{}.
This property also greatly simplifies the counting of writes in our simulator and makes the optimizations to reduce writes easier.


\myparagraph{\SPLIT{}$(T,k)\to (T_L,T_R)$}: \SPLIT{} can be viewed as the inverse of \JOIN{} that takes a balanced BST and a key $k$, and splits the tree into two smaller trees $T_L$ that contains keys smaller than key $k$, and  $T_R$ for keys larger than key $k$.
\SPLIT{} can be trivially implemented using \JOIN{} by a recursive approach.

\myparagraph{Bulk Operations.} Using \JOIN{} as a black box, either single or bulk updates can be implemented simply and generally across different balancing schemes.
As an example, we give the algorithm of taking the union of two BSTs (or sets/maps) in Algorithm~\ref{algo:join}.
The algorithm is based on divide-and-conquer: $T_1$'s root is used to partition all elements in two trees into two disjoint set, one with $T_1$'s left subtree and $T_l$, and the other with $T_1$'s right subtree and $T_r$.
Then we union the two subproblems recursively and independently, and join the two output trees using $T_1$'s root.
The correctness and running time are shown in~\cite{blelloch2016just}.

\medskip

There are several benefits of using this framework for our implementation and experiment.
First, as we just explained, different updates have the uniform code on different types of BSTs (except for the \JOIN{}), which justifies the performance by the different balancing criteria for the BSTs, instead of the different implementations for different trees.
Second, although the joined-based implementation operates on two trees, one can check that when one tree contains only a singleton element, the algorithm runs the same as the algorithm of a single insertion on each type of the BSTs.
The deletion can also be implemented by taking the difference by the original tree and a tree with a single element.
As a result, the joined-based implementation is strictly more powerful.
Lastly, we can also run interesting experiments on more operations like bulk updates, and compare the results on different BSTs.

\begin{algorithm}[t]
\caption{The union function}
\label{algo:join}
    \myfunc{\upshape\textsc{Union}($T_1$, $T_2$)} {
        \lIf {$T_1 = \textsc{Nil}$} {
            \Return $T_2$
        }
        \lIf {$T_2 = \textsc{Nil}$} {
            \Return $T_1$
        }
        $\langle T_l, T_r \rangle \gets$ \SPLIT{}($T_2$, $T_1$'s root)\\
        \Return \JOIN($T_1$'s root, \textsc{Union}($T_1$'s left subtree, $T_l$), \textsc{Union}($T_1$'s right subtree, $T_r$))
    }
\end{algorithm}

\subsubsection{Experiments}

In this section, we show our experimental results on the counts of read/write transfers for different settings. 
Due to the page limit, our experiment mainly focuses on the performance of various binary search trees (AVL trees, red-black trees, and treaps) with different batch sizes.

In the experiment, we first insert $m=10^6$ million integers as keys to a tree $T$ (empty at the beginning), drawn from a uniform distribution from 32-bit unsigned integers, and then delete them in a uniformly random order.
The insertion and deletion are grouped in batches of size $s$, indicating that the insertions are $\lceil m/s \rceil$ unions on the main tree $T$ with trees of size $s$.
The deletions are also batched in $\lceil m/s \rceil$ bulks of size $s$.
In our experiment, we construct a smaller tree for each batch and then call the union or difference function we just mentioned.
We note that if all update elements are given in advance, we can also sort them and put them in a list.
Since this is a more specific case, our experiment is based on the tree-tree updates.

The node size in all different types of tress is 16 bytes.
Each tree node stores four 4-byte data blocks to hold the key, the left and right pointers, and the balancing information.
The cache contains 10,000 cache-lines, similar to the setting for unordered sets.

Table \ref{tab:bst-table} shows the experimental results on BSTs with different balancing schemes with various batch sizes.
The numbers are read and write transfers per update.

\begin{table}[th]
\def\arraystretch{1.2}
  \centering
  \small
    \begin{tabular}{r|rr|rr|rr|r@{ }@{ }@{ }r@{ }@{ }r|r@{ }@{ }@{ }r@{ }@{ }r}
    \multicolumn{13}{c}{\textbf{(a) Insertion / Union}} \\
    \toprule
    \multicolumn{1}{c|}{\textbf{Batch}}  & \multicolumn{2}{c|}{\textbf{AVL}} & \multicolumn{2}{c|}{\textbf{RB-Tree}} & \multicolumn{2}{c|}{\textbf{Treap}}  & \multicolumn{3}{c|}{$\bm{\wcost{}=10}$} & \multicolumn{3}{c}{$\bm{\wcost{}=100}$} \\
    \cline{2-13}
    \multicolumn{1}{c|}{\textbf{Size}}      & \multicolumn{1}{c}{\textbf{RT}} & \multicolumn{1}{c|}{\textbf{WT}} & \multicolumn{1}{c}{\textbf{RT}} & \multicolumn{1}{c|}{\textbf{WT}} & \multicolumn{1}{c}{\textbf{RT}} & \multicolumn{1}{c|}{\textbf{WT}} &
    \multicolumn{1}{c}{\textbf{AVL}} & \multicolumn{1}{@{ }c@{}}{\textbf{RB-T}} & \multicolumn{1}{@{ }c|}{\textbf{Treap}} & \multicolumn{1}{c}{\textbf{AVL}} & \multicolumn{1}{@{ }c@{}}{\textbf{RB-T}} & \multicolumn{1}{@{ }c}{\textbf{Treap}} \\
    \hline
    \textbf{1} & 11.28 & 2.83  & 12.00 & 3.48  & 16.61 & 1.79  & 39.6  & 46.8  & 34.5  & 295   & 360   & 196 \\
    \textbf{1k} & 12.67 & 2.89  & 13.29 & 3.66  & 17.18 & 1.89  & 41.6  & 49.9  & 36.1  & 302   & 379   & 206 \\
    \textbf{10k} & 6.18  & 2.68  & 6.40  & 3.01  & 7.33  & 1.85  & 33.0  & 36.5  & 25.8  & 274   & 308   & 192 \\
    \textbf{100k} & 2.54  & 1.84  & 2.54  & 1.86  & 2.56  & 1.66  & 20.9  & 21.2  & 19.2  & 186   & 189   & 169 \\
    \bottomrule
    \end{tabular}
    \begin{tabular}{r|rr|rr|rr|r@{ }@{ }@{ }r@{ }@{ }r|r@{ }@{ }@{ }r@{ }@{ }r}
    \multicolumn{13}{c}{\textbf{(b) Deletion / Difference}} \\
    \toprule
    \multicolumn{1}{c|}{\textbf{Batch}} & \multicolumn{2}{c|}{\textbf{AVL}} & \multicolumn{2}{c|}{\textbf{RB-Tree}} & \multicolumn{2}{c|}{\textbf{Treap}}   & \multicolumn{3}{c|}{$\bm{\wcost{}=10}$} & \multicolumn{3}{c}{$\bm{\wcost{}=100}$} \\
    \cline{2-13}
    \multicolumn{1}{c|}{\textbf{Size}}  & \multicolumn{1}{c}{\textbf{RT}} & \multicolumn{1}{c|}{\textbf{WT}} & \multicolumn{1}{c}{\textbf{RT}} & \multicolumn{1}{c|}{\textbf{WT}} & \multicolumn{1}{c}{\textbf{RT}} & \multicolumn{1}{c|}{\textbf{WT}} &
    \multicolumn{1}{c}{\textbf{AVL}} & \multicolumn{1}{@{ }c@{}}{\textbf{RB-T}} & \multicolumn{1}{@{ }c|}{\textbf{Treap}} & \multicolumn{1}{c}{\textbf{AVL}} & \multicolumn{1}{@{ }c@{}}{\textbf{RB-T}} & \multicolumn{1}{@{ }c}{\textbf{Treap}} \\
    \hline
    \textbf{1} & 13.83 & 2.72  & 16.17 & 5.17  & 17.85 & 1.98  & 41.1  & 67.8  & 37.7  & 286   & 533   & 216 \\
    \textbf{1k} & 15.11 & 2.79  & 17.65 & 5.21  & 18.52 & 2.08  & 43.0  & 69.8  & 39.3  & 294   & 539   & 227 \\
    \textbf{10k} & 8.17  & 2.69  & 10.43 & 3.44  & 9.09  & 2.06  & 35.1  & 44.9  & 29.7  & 278   & 355   & 215 \\
    \textbf{100k} & 3.22  & 1.99  & 3.54  & 2.43  & 3.23  & 1.78  & 23.2  & 27.8  & 21.1  & 203   & 246   & 182 \\
    \bottomrule
    \end{tabular}
    \begin{tabular}{p{2cm}<{\centering}|p{2cm}<{\centering}|p{2cm}<{\centering}}
    \multicolumn{3}{c}{\textbf{(c) Average Tree Depth}} \\
    \toprule
           {\textbf{AVL}} & {\textbf{RB-Tree}} & {\textbf{Treap}} \\
\hline
           {19.39} & {19.62} & {26.48} \\
    \bottomrule
    \end{tabular}%
    \caption{Numbers of \rwt{s} and asymmetric I/O costs of different BSTs with various batch sizes.  The numbers are divided by $10^6$ (i.e., per inserted/deleted elements).  The write-read ratio $\wcost{}$ are selected to be typical projected values 10 (latency, bandwidth) and 100 (energy). }
      \label{tab:bst-table}
\end{table}
\begin{figure*}[ht]
\begin{center}
    \begin{minipage}[t]{0.4\textwidth}
        \includegraphics[width=\textwidth]{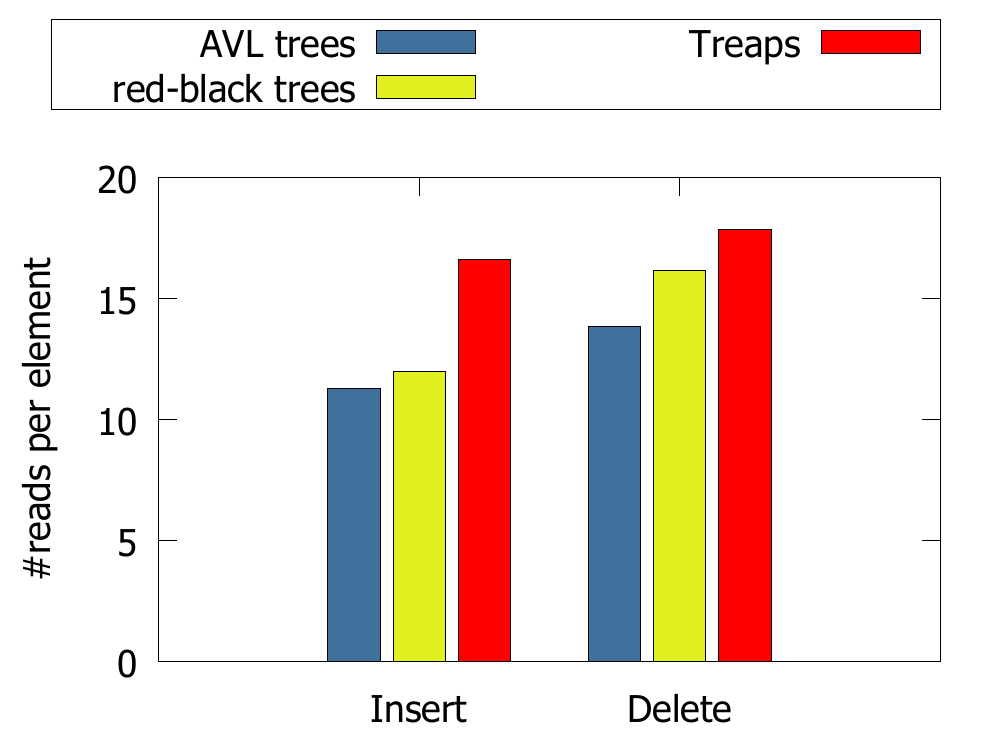}\\ \centering{\small (a) Number of reads}
    \end{minipage}~~~
    \begin{minipage}[t]{0.4\textwidth}
        \includegraphics[width=\textwidth]{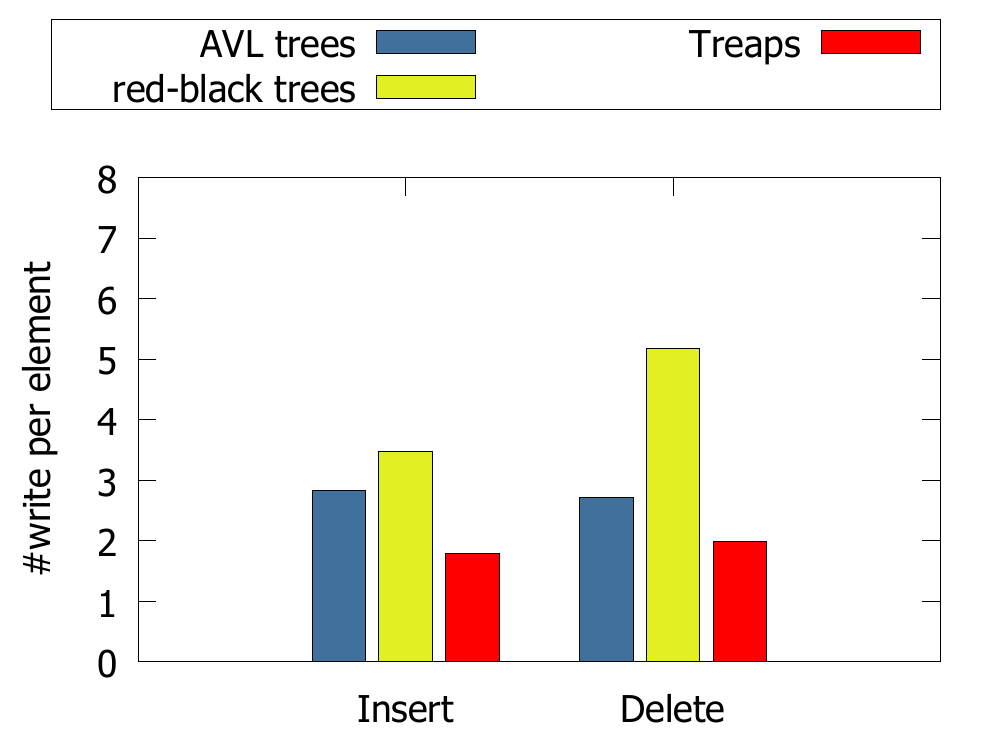}\\ \centering{\small (b) Number of writes}
    \end{minipage}
    \captionof{figure}{The number of \rwt{s} of different BSTs on single insertion/deletion.  Data are from the first rows in Table~\ref{tab:bst-table}(a) and \ref{tab:bst-table}(b).}\label{fig:bst-rw}
\end{center}
\begin{center}
    \begin{minipage}[t]{0.4\textwidth}
        \includegraphics[width=\textwidth]{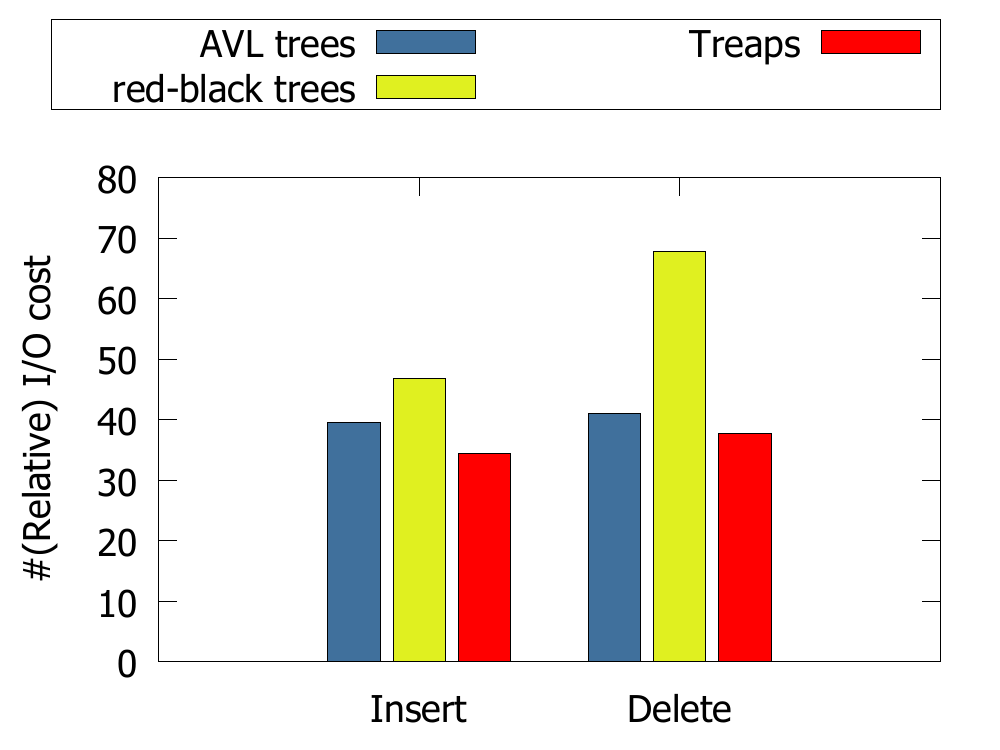}\\ \centering{\small (a) $\wcost=10$}
    \end{minipage}~~~
    \begin{minipage}[t]{0.4\textwidth}
        \includegraphics[width=\textwidth]{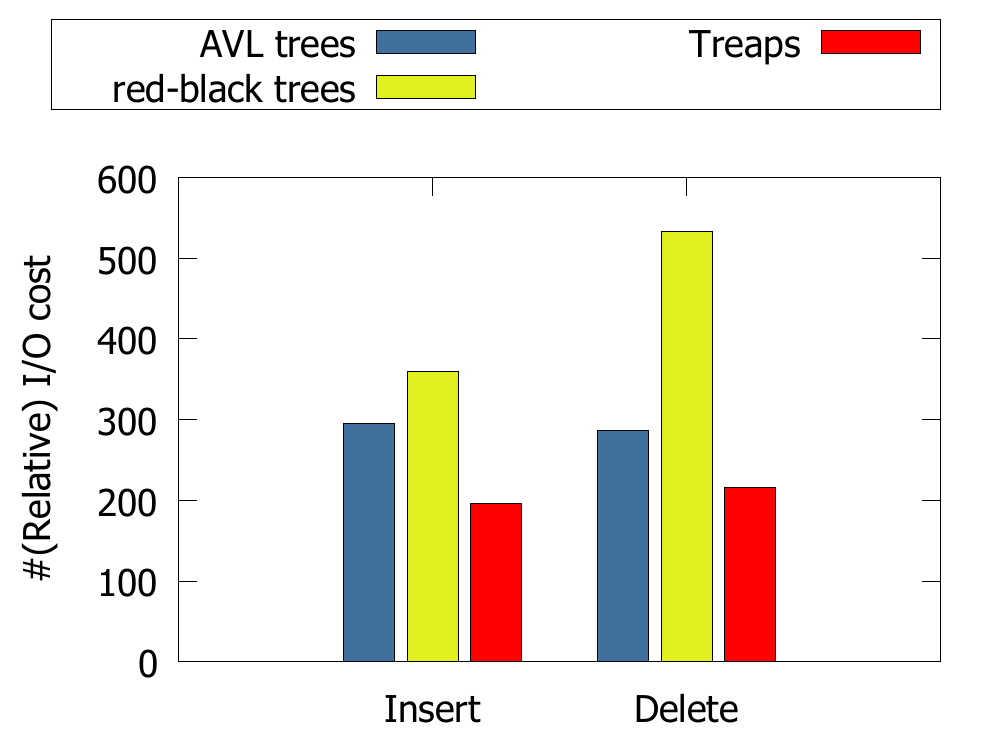}\\ \centering{\small (b) $\wcost=100$}
    \end{minipage}
    \captionof{figure}{The I/O cost of different BSTs on single insertion/deletion.  Data are from the first rows in Table~\ref{tab:bst-table}(a) and \ref{tab:bst-table}(b).}\label{fig:bst-io}.
\end{center}
\end{figure*}%

\myparagraph{Batch size 1.}

We first look at the case corresponding the single insertions and deletions, and the results are shown in Figure~\ref{fig:bst-rw} and~\ref{fig:bst-io}.
Regarding the number of writes, treaps show the best performance.
This is easy to understand since treaps do not modify any information for rebalancing during insertions/deletions.
The structure of treaps is deterministic once the priorities are decided.
The priorities are set before the merging (deleting), and never changed later.
For such reasons, treaps require much fewer writes per update compared to AVL and red-black trees.

The AVL and red-black trees maintain the balancing information on each tree node that needs to be updated when the subtree height changes.
Then more writes are used due to these updates.
Between these two types of BSTs, red-black trees require more writes, since there are also some color flips on the siblings of the updated tree path.
Such flips are extremely cheap in the classic symmetric setting but will cost much in the asymmetric setting when writes become expensive.

The fewer writes for treaps come together with the extra cost on more reads.
Treaps are less strictly balanced compared to the other two trees, which saves the writes to maintaining such balancing, but leads to larger average tree depth (shown in Table~\ref{tab:bst-table}(c), about 30\% deeper than the other trees).
The average depths for AVL and red-black trees are close to optimal, which is 18.96 for a perfectly balanced tree with $10^6$ nodes.
Because of the shallower average depth, the number of reads required in either updates or queries is much small on these two trees.

\myparagraph{Larger batch sizes.}

We now discuss how does the batch size affect the numbers of read and write transfers.
The numbers are given in Table~\ref{tab:bst-table}.

When the batch size increases but remains small, the reads and writes almost remain the same and slightly increase.
When the batch size is smaller, the elements in a batch and the paths to visit them are always loaded into the cache once and always stay there.
Therefore, the different batch sizes less than 100 do not affect the performance much.

The peak I/O cost per update is around batch size 1000, as we show here.
In such case, the overall footprint of a batch update no longer fit into the cache, which may lead to two loads per tree node (once in the split phase and the other in the join phase).

However, the cost turns down when the batch size grows over 1000.
The reason is that, the number of tree nodes visited during each bulk update is $\Theta(s\log (m/s))$ (recall that $s$ is the bulk size), which is also the time complexity~\cite{blelloch2016just} of this process.
Namely, we need to visit $O(\log (m/s))$ tree nodes per inserted node, so we touch fewer nodes as $s$ goes up.
Compared to the single updates, each node on multiple tree paths\footnote{Usually on the top part of the tree.  For example, every single insertion visits the root node.} is only looked at and compared with once in the joined-based bulk updates, and this is from which the improvement comes.
Since the top levels in the trees are visited more frequently, they usually stay in the cache.
As a result, the saved memory accesses for small batch size are hidden by the function of the cache.
However, when the batch size keeps growing and exceed the cache size, then the saved memory accesses lead to lower reads, as shown in Table~\ref{tab:bst-table}.

The number of writes also decreases for larger batch sizes in AVL and red-black trees, because of the previously stated reason.
When the elements are inserted or deleted one by one, the balancing factor of a node on by multiple tree paths can be updated multiple times.
On the other hand, when the tree is updated in a bulk, such information will be updated by at most once at the join point, which saves the number of writes to the asymmetric memory.
However, since treaps do not need to update such information, we cannot observe a significant drop-off on writes for larger batch sizes.


\myparagraph{Queries.}

We have not explicitly tested the I/O costs for queries, since most queries (like finding or checking a key, locating the $k$-th element) have the same memory access pattern as the insertions.
The only difference is that they do not modify the tree, so there will be no writes.
These updates may flush the dirty cache lines, and thus slightly increase the writes.
Our experiment shows that if we have the same number of queries and insertions, the number of writes increases by no more than 5\%, which is insignificant.
Therefore, we believe that we can ignore such changes in the most cases.

\subsubsection{Conclusions}

In this section, we theoretically analyze the asymmetric I/O costs of different types of binary search trees.
We show that red-black trees, the insertions for AVL trees, and treaps on expectation have an optimal asymptotic cost ($\Theta(\wcost{}+\log n)$ per update).

We then test the actual performance by conducting experiments based on the join-based implementation, and show that treaps have the best update cost in most cases.
The advantage comes from a looser balancing constraint, which also leads to a larger tree depth and query costs.
As a result, AVL tree will be a better option if the queries are much more than the updates.

\section{Sorting}
\label{sec:sort}

Sorting is one of the most fundamental algorithms and building blocks in algorithm design and programming.
In this section, we analyze, performance engineer and experiment the performance of the existing sorting algorithms in the asymmetric setting, which include quicksort, mergesort, BST sort and samplesort.

\subsection{Algorithms and Implementations}
\label{sec:sort-analysis}

We discuss four sorting algorithms, which are either used as the state-of-the-art implementations or have efficient theoretical guarantees on asymmetric I/O cost.
All four algorithms requires $O(n\log n)$ (optimal) comparisons.

\myparagraph{Quicksort.}  Quicksort is one of the commonly-used sorting algorithms in both sequential and parallel setting.
It picks a random pivot (or the median among several random samples) that partitions the array into three contiguous subsets containing the input values that are less than, equal to, and larger than the pivot respectively, based on a scan-based approach.
Then two recursive calls are made to the less-than and larger-than subsets, until the base case (we set it to be 16 elements) is reach and the algorithm switch to an insertion sort.  In our implementation, the pivot is selected to be the median of three random positions.

We now analyze the I/O cost of quicksort.
Once the subproblem size is no more than $M$, the \smallmem{} size, the subproblem will fit into the cache and no more read and write transfers are required for this subtask thereafter.
The I/O cost $Q$ is thus $O(\wcost{}n/B\cdot \log (n/M))$, where the number of memory transfers to partition the array in the same recursive level add up to $O(n/B)$ and w.h.p.\ the algorithm requires $O(\log (n/M))$ levels to reach the subtask with size no more than $M$.

\myparagraph{Mergesort.}  Mergesort is another textbook sorting algorithm that is stable and easy to parallel.
The implement of mergesort has different versions, and here we discuss an I/O-efficient version.
The algorithm partitions the array into two equal-length parts and recursively sort every subproblem individually.
After the computation of both subtasks is finished, a scan-based process merges the two sorted subsequences into the final output.
To implement the algorithm efficiently, we use the \rotating{} so that every element is moved only once in one merge process.
An extra round of data copy is applied if the latest merge leaves the sorted result in the temporal array.

Similarly to quicksort, no further memory transfers are used in subproblems with size no more than $M$.  Therefore the I/O cost $Q$ of mergesort is also $O(\wcost{}n/B\cdot \log (n/M))$.  More specifically, $Q = (1+\wcost{})n/B\cdot \lceil\log_2(n/M) \rceil$ when $\lceil\log_2(n/M) \rceil$ is even, and otherwise there is an extra $(1+\wcost{})n/B$.

\myparagraph{BST sort.}  BST sort treat the input as an ordered set and all elements are maintained in a binary search tree.
Then the in-order traversal of the tree is the sorted output.  The BST can either be balanced, or does not apply any rotation but the elements are inserted in a random order.
Both cases give I/O cost of $O(n\log n+n\wcost{})$ on $n$ input elements.
It is typically used when the input is adaptive: we can insert and delete elements dynamically, while the sorted result are maintained at any time.
In this case, some balancing schemes are required if the insertions and deletions are biased.

In our implementation elements are inserted in a random order, which is done in two phases: the first phase generates $n$ uniformly random integers $a_1,\cdots, a_n$ between 1 to $n$ using some hash functions, and inserting the $a_i$-th elements into the tree with \emph{no} rotations (check whether this element is already inserted before the actual insertion); then in the second phase we insert every uninserted element into the tree in the input order.  This guarantees that w.h.p.\ the gap between the elements inserted in the first phase is $O(\log n)$ in the output order.  Hence the tree depth is $O(\log n)$ disregarding any input order, and thus inserting one element requires $O(\log n)$ reads and two random writes (the boolean flag that this node is inserted, and to change the parent's pointer).

\myparagraph{Samplesort.}  Samplesort is a class of sorting algorithms that are based on divide-and-conquer paradigm with multiple pivots and especially commonly-used in the multicore setting.  To minimize the I/O cost and number of memory transfers, in this paper, we discuss a cache-aware implementation of samplesort.

Given the cache size $M$ and block size $B$, if the input size is smaller than $M$ then we simply call quicksort.  Otherwise, we pick $M$ random samples from the input, sort the samples, and pick the $B$-th, $2B$-th, ... , $(M-B)$-th samples as the \textbf{pivots}\footnote{\small If $M<n/B$ then we pick $n/B$ samples, and this will happen only once w.h.p.\ in the algorithm.  In practice we replace $M$ to be $M/c'$ for some small constant $c'>1$ to ensure everything that should be in the cache is actually in the cache.} and they form $M/B$ buckets.  Then for each input element, we binary search (an implicit search tree) its associated bucket.  Finally, we again samplesort the elements in each bucket individually and recursively.

The algorithm partition the input into $M/B$ almost equal-size subproblems with $O(n/B)$ read and write transfers.  The I/O cost of this algorithm on average is $\displaystyle O\left({\wcost{}n\over B}\log_{M\over B}{n\over B}\right)$.

The number of \wrt{} of samplesort in our implementation is highly optimized so that one round of samplesort only requires three sequential reads and one sequential write per input element.
After picking the pivots, we first loop over the input to binary search the associate bucket of each element.
However, we do not store this value; instead, we only modify the counters of each bucket.
After that, we have known the number of elements in each bucket, and we apply a prefix sum compute the offset of each element.
Finally, all elements are distributed to their associated buckets based on the offsets.
The algorithm only requires three reads and one write for each element, and all other operations are all within the cache.
Notice that after one round, the data are stored in another array, so the final results will be moved back to the original array if they happen to be in the other one.

\begin{table}[t]
\small
\def\arraystretch{1.2}
\centering
\begin{tabular}{p{2cm}p{4cm}<{\centering}}
\toprule
  \quad Algorithm & I/O cost \\
  \midrule
  \quad Quicksort & $O(\wcost n/B\cdot \log (n/M))^*$\\
  \quad Mergesort & $O(\wcost n/B\cdot \log (n/M))$\\
  \quad BST sort & $O(n\log n+n\wcost{}))$\\
  \quad Samplesort & $O({\wcost{}n/B}\cdot\log_{M/B}(n/B))^*$\\
\bottomrule
\end{tabular}
\caption{List of I/O costs on sorting algorithms.  (*) indicates the bounds hold with high probability ($1-n^{-c}$ for arbitrary $c>0$).}
\end{table}

\smallskip There do exist other sorting algorithms like heapsort, shellsort, bitonic sort, etc.  Their performances regarding I/O cost however, are less competitive against the previous sorting algorithms due to either more work or inefficient memory-access pattern.
We did not compare with previous work in~\cite{Viglas14,BFGGS15} in this paper, since they are not optimal in terms of comparisons and have tunable parameters in the algorithm, that makes the comparison among them inconclusive.
Instead as a very first paper on this topic, we focus on simple algorithms and draw interesting conclusions that can also be useful in implementing these more complicated approaches in the future.

\begin{figure*}[p]
\centering
\def\arraystretch{1.2}
\small
\begin{tabular}{c|rrr|rrr|rrr|rrr}
\toprule
     \textbf{Cache} &      \multicolumn{ 3}{c|}{\textbf{Quicksort}} &      \multicolumn{3}{c|}{\textbf{Mergesort}} &        \multicolumn{ 3}{c|}{\textbf{BST sort}} &     \multicolumn{3}{c}{\textbf{Samplesort}} \\
     \cline{2-13}

      \textbf{Size} &         \textbf{RT} &         \textbf{WT} &      \textbf{\#comp} &         \textbf{RT} &         \textbf{WT} &      \textbf{\#comp} &         \textbf{RT} &         \textbf{WT} &      \textbf{\#comp} &         \textbf{RT} &         \textbf{WT} &      \textbf{\#comp} \\
\hline
       \textbf{100} &       2.06 &       2.04 &      39.49 &       4.01 &       2.00 &      24.11 &      24.49 &       2.04 &      43.76 &       1.60 &       0.82 &      63.05 \\

      \textbf{1000} &       1.58 &       1.57 &      39.49 &       3.11 &       1.56 &      24.11 &      19.39 &       2.04 &      43.76 &       1.01 &       0.51 &      56.63 \\

     \textbf{10000} &       1.11 &       1.10 &      39.49 &       2.25 &       1.13 &      24.11 &      14.17 &       2.03 &      43.76 &       0.50 &       0.25 &      52.47 \\
\bottomrule
\end{tabular}
\captionof{table}{Number of read and write transfers of different sorting algorithms on $10^7$ random i.i.d.\ double-precision float-point numbers.  Values in the table are numbers of read and write transfers and comparisons per input element.  Cache size is measured by the number of 64-byte cache-lines.}
\label{tbl:sort}

\bigskip
\begin{tabular}{c|rrr|rrr|rrr|rrr}
\toprule
     \textbf{Cache} &      \multicolumn{ 3}{c|}{\textbf{Quicksort}} &      \multicolumn{3}{c|}{\textbf{Mergesort}} &        \multicolumn{ 3}{c|}{\textbf{BST sort}} &     \multicolumn{3}{c}{\textbf{Samplesort}} \\
     \cline{2-13}

      \textbf{Size} &         \textbf{RT} &         \textbf{WT} &      \textbf{\#comp} &         \textbf{RT} &         \textbf{WT} &      \textbf{\#comp} &         \textbf{RT} &         \textbf{WT} &      \textbf{\#comp} &         \textbf{RT} &         \textbf{WT} &      \textbf{\#comp} \\
\hline
       \textbf{100} &      19.83 &       1.14 &      37.68 &      17.04 &       1.02 &      23.30 &      46.24 &       1.93 &      40.93 &       6.19 &       0.79 &      35.97 \\

      \textbf{1000} &      15.57 &       0.91 &      37.68 &      13.42 &       0.81 &      23.30 &      35.30 &       1.92 &      40.93 &       3.93 &       0.45 &      34.34 \\

     \textbf{10000} &      11.12 &       0.67 &      37.68 &       8.96 &       0.56 &      23.30 &      24.65 &       1.90 &      40.93 &       2.38 &       0.38 &      34.43 \\
\bottomrule
\end{tabular}
\caption{Number of read and write transfers of different sorting algorithms on $2\times10^6$ indices pointing to structures with average size of 64 bytes.  Other setup is the same as that in Table~\ref{tbl:sort}.}
\label{tbl:sort-ind}

\begin{center}
    \begin{minipage}[t]{0.45\textwidth}
        \includegraphics[width=\textwidth]{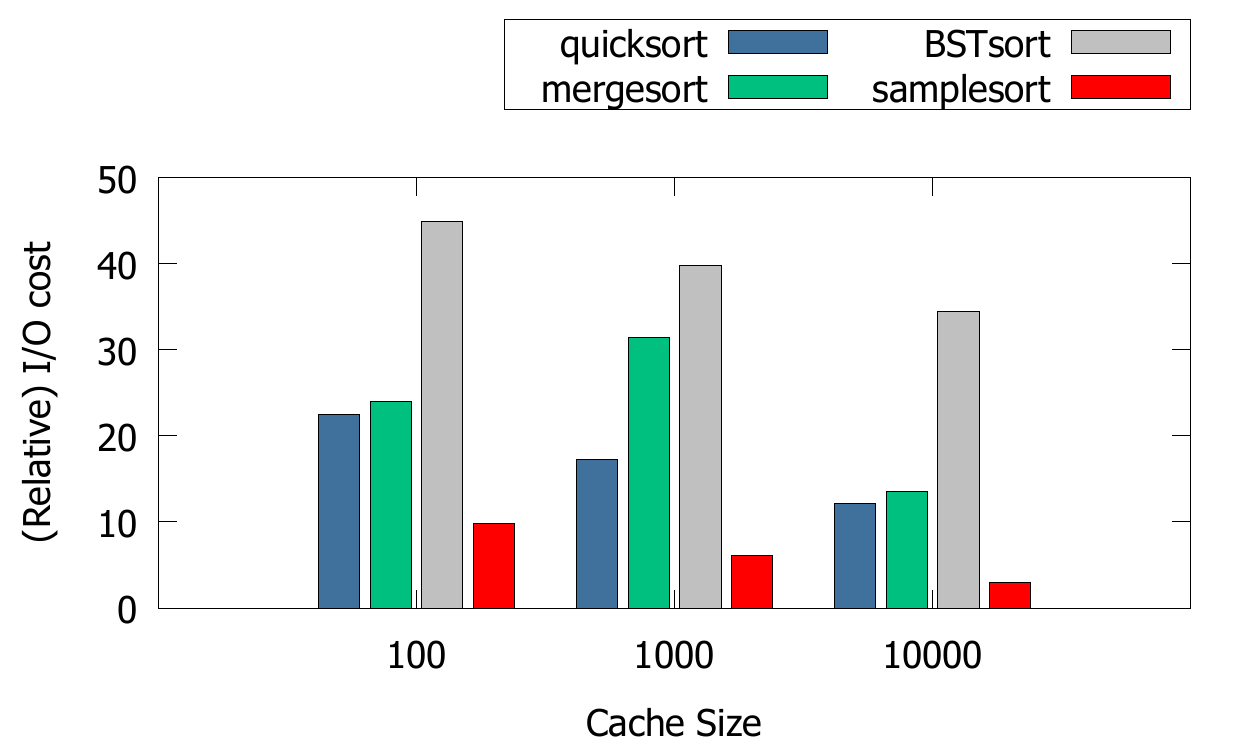}\\ \centering{\small (a) $\wcost=10$}
    \end{minipage}~~~
    \begin{minipage}[t]{0.45\textwidth}
        \includegraphics[width=\textwidth]{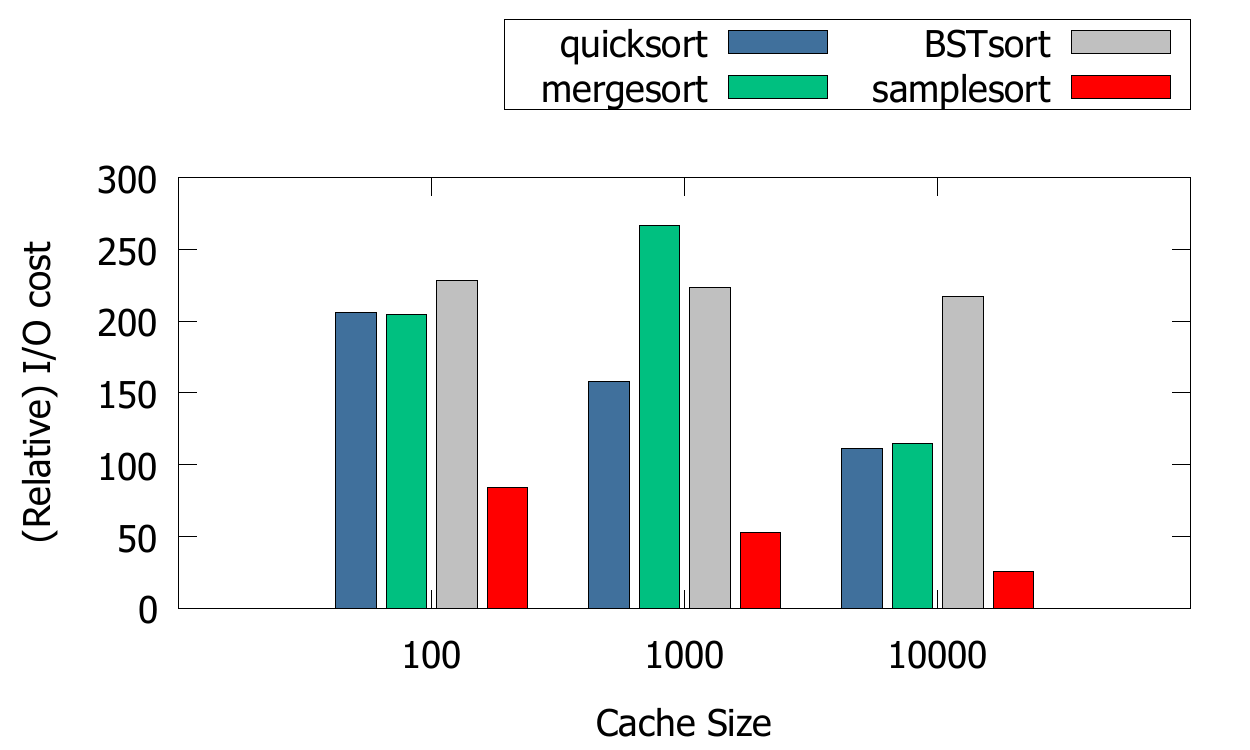}\\ \centering{\small (b) $\wcost=100$}
    \end{minipage}
    \captionof{figure}{I/O costs of different sorting algorithms on $10^7$ \textbf{doubles} with various read-write asymmetry $\wcost$.  Numbers are from Table~\ref{tbl:sort}.}\label{fig:addingedges}
\end{center}
\begin{center}
    \begin{minipage}[t]{0.45\textwidth}
        \includegraphics[width=\textwidth]{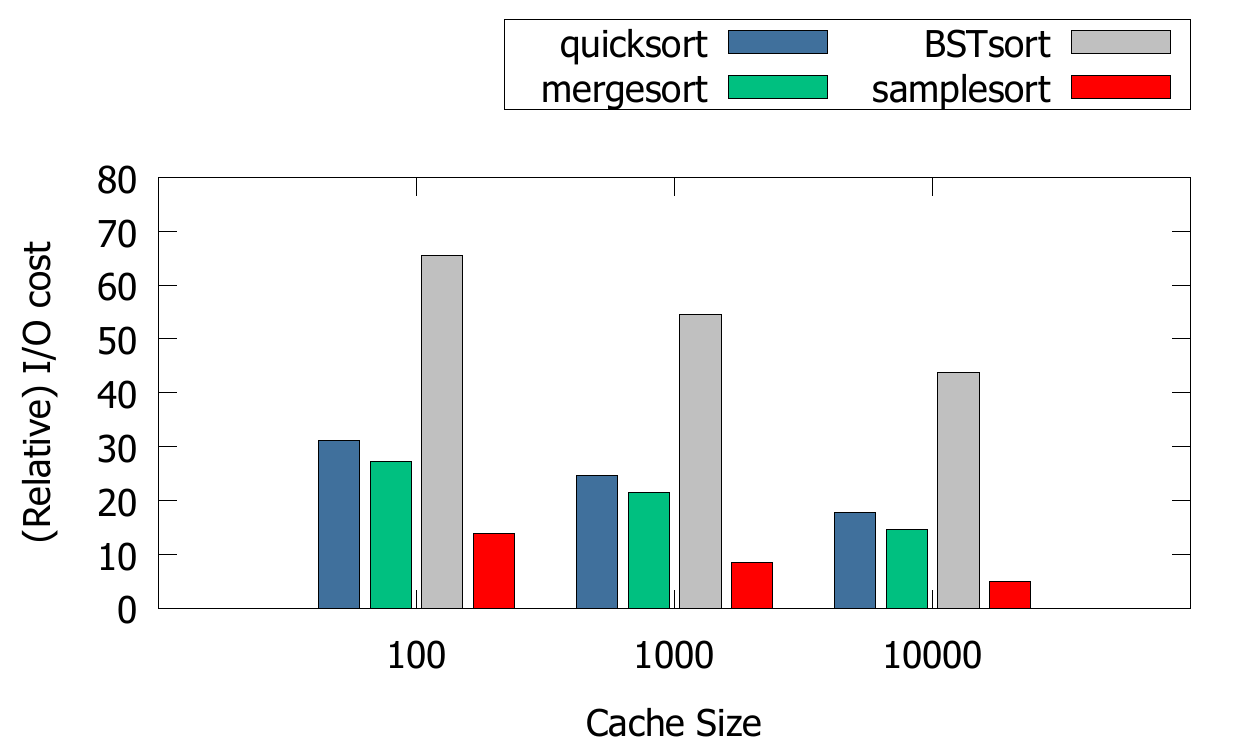}\\ \centering{\small (a) $\wcost=10$}
    \end{minipage}~~~
    \begin{minipage}[t]{0.45\textwidth}
        \includegraphics[width=\textwidth]{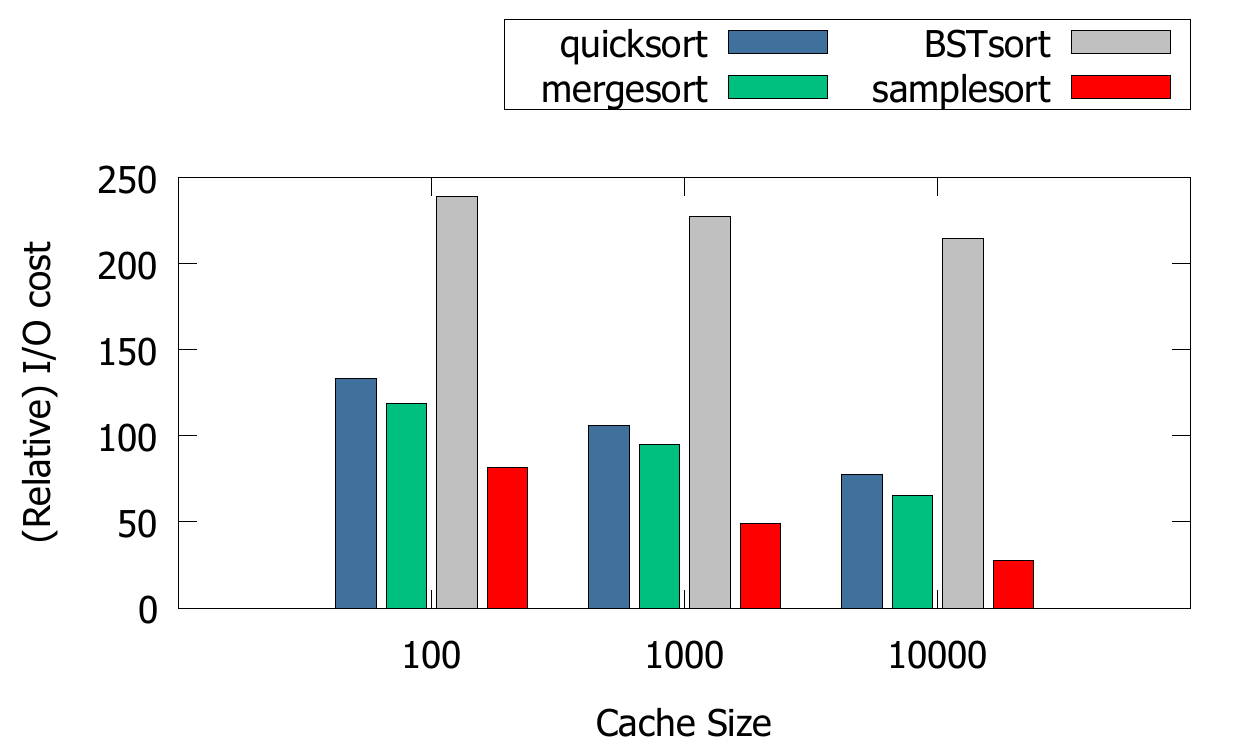}\\ \centering{\small (b) $\wcost=100$}
    \end{minipage}
    \captionof{figure}{I/O costs of different sorting algorithms on $2\times 10^6$ \textbf{pointers} with various read-write asymmetry $\wcost$.  Numbers are from Table~\ref{tbl:sort-ind}.}\label{fig:sort-doubles}
\end{center}
\vspace{-2em}
\end{figure*}

\subsection{Experiments}

In the experiment we test our implementations of the four sorting algorithms on the numbers of \rwt{s} with various cache sizes.
We set the input into two categories: one that we indeed move the data, and the other that the algorithm sorts the indirect pointers pointing to the data fields.
The first case is when we are sorting integers, real numbers or any structures with small and fixed size like graph edges, key-value pairs of integers and/or pointers, etc.
The second case is when the input is irregular, like strings, texts, images, or any structure that is either large or the sizes vary among different elements.
In this case, moving the data is expensive, so we will sort and output a list of indices pointing to the actual data fields.  Both cases are widely used in practice.


\myparagraph{Performance on sorting float-point numbers.}
We now show the experiments that perform data movement.  We sort 10 million random i.i.d.\ double-precision float-point numbers and in this case the block size $B=8$.  The numbers of comparisons and \rwt{s} of four algorithms are summarized in Table~\ref{tbl:sort}.  The I/O costs of the algorithms for two typical values of $\wcost{}$ (10 and 100) are visualized in Figure~\ref{fig:sort-doubles}.

Both quicksort and mergesort require about $\log_2 (n/M)$ rounds, which is 8 to 16, to fully fit the subproblems into the \smallmem{}.  Both algorithms require approximately the same number of writes, since quicksort is in-place while mergesort is not and has double memory footprint, but the partition of quicksort is not exactly even.  These factors offset each other.  However, the \cm{} are doubled in mergesort, and in practice causing its less efficient practical performance.  Mergesort requires fewer comparisons, but this does not lead to a significant difference in modern architecture, compared to the I/O cost to the main memory.

BST sort, as we analyzed theoretically, requires about $\log_2(nB/M)$ reads and two random writes plus some small cost on initializations.  BST sort uses less writes compared to quicksort or mergesort when $2B<\log_2(n/M)$, which is hard to be satisfied given the parameters of the hardware.
BST sort thus will still be less efficient to sort small and fixed-structure entries in future memories.

The experiment result of samplesort follows our analysis as well: each round of sample sort approximately requires three sequential reads and one sequential write, and the algorithm uses roughly $\lceil\log_{M/B}(n/B)\rceil$ rounds to reach the base case.  Other costs (sampling and sorting pivots, transposing the offsets, etc.) are negligible.

Based on our previous experience, after highly optimizing the performance in the sequential setting, samplesort is slightly slower than quicksort (but much faster in multicore parallelism) even though the I/O cost of samplesort is lower.  We believe the reason to be that, samplesort does not explicitly utilize L1 and L2 caches (the binary search, offset transpose, etc.) so most \smallmem{} accesses are toward L3 cache.
The memory access pattern for quicksort, however, is mostly sequential scanning, which is highly optimized (like by prefetching) in the current architecture.
Since the write costs will significantly increase for future non-volatile main memory, we project that samplesort will be more efficient in this setting, since the I/O cost will very likely be the bottleneck of all these sorting algorithms (similar to the existing multicore parallel setting).

\myparagraph{Sorting larger entries.}
As shown previously, although BST sort requires only at most random two writes per insertion, this cost is still significant compared to other sorting algorithms that have better spatial locality in memory access.
However, when the size of each input entry becomes larger, each cache-line holds fewer entries and the number of \wrt{s} per entry increases.

We run the experiment with a fixed number of inputs ($10^7$) and cache-lines (1000), and varies the size of each input entry from 8 bytes to 64 bytes.
The results are shown in Table~\ref{tab:sort-vary-size}.
Based on the analysis in Section~\ref{sec:sort-analysis}, for quicksort, mergesort, and samplesort, the I/O cost is proportional to the entry size.
However, BST sort is mostly not affected.
The number of reads just marginally increases since the cache can hold less top-tree levels, and slightly more writes are required since the overall footprint of the search tree increases as the growth of the entry size.
From the results, we can see that when the entry size becomes at least 64 bytes, BST sort outperforms all other algorithms in the asymmetric setting.

\myparagraph{Performance on sorting via pointers.}  In this experiment, we sort 2 million random strings that the characters are stored contiguously in the memory with an average size of 64 bytes.  The input also contains 2 million 8-byte pointers to the strings, and we only sort the pointers.

The implementations of quicksort, mergesort, and BST sort are identical except that each comparison requires two indirect addresses to locate the data field.  Samplesort, since it is cache-aware, requires two modifications: first, no oversampling for the pivots since now each pivot requires one or two cache-lines for its data; second, instead binary-search the bucket of each element twice, we now store the bucket label after the first search to avoid the second search (roughly double the writes while half the reads).

The numbers of \rwt{s} are reported in Table~\ref{tbl:sort-ind} and visualized for two typical values of $\wcost{}$ (10 and 100) in Figure~\ref{fig:sort-doubles}.
Writes of quicksort, mergesort, and BST sort stay approximately the same as the previous experiment (the decrease in values is due to fewer input elements).  Reads are now increased by about $\log_2(nB/M)$, the cost of locating the data fields.

For each round in sample sort, each element now requires one random read, three sequential reads and two sequential writes (i.e.\ 1.375 \cm{s} and 0.25 \wrt).
Also, one extra random read per element is in the base case to load the data into the cache.
The algorithm requires $O(\log_{M/B}(n))$ rounds to reach the base case, and in the experiment, this round number is approximately 3, 2 and 1 for three cache sizes.
Finally, output needs to be moved back when reaching the base case in an odd number of rounds.

As what we understand, the wall-clock performance of samplesort when sorting pointers on current architecture is already faster than the other sorts.
This is because samplesort requires fewer I/Os and comparisons, and all algorithms randomly access the memory so techniques like prefetching cannot help.  This gap will be even enlarged in future NVMs when the write costs being exaggerated.

\begin{table}[t]
  \centering
\def\arraystretch{1.2}
\small
    \begin{tabular}{c|r@{  }@{  }rr@{  }@{  }rr@{  }@{  }rr@{  }@{  }r}
    \toprule
    \textbf{Entry} & \multicolumn{2}{c}{\textbf{Quicksort}} & \multicolumn{2}{c}{\textbf{Mergesort}} & \multicolumn{2}{c}{\textbf{BSTsort}} & \multicolumn{2}{c}{\textbf{Samplesort}} \\
    \cline{2-9}
    \textbf{Size} & \multicolumn{1}{c}{\textbf{RT}} & \multicolumn{1}{c}{\textbf{WT}} & \multicolumn{1}{c}{\textbf{RT}} & \multicolumn{1}{c}{\textbf{WT}} & \multicolumn{1}{c}{\textbf{RT}} & \multicolumn{1}{c}{\textbf{WT}} & \multicolumn{1}{c}{\textbf{RT}} & \multicolumn{1}{c}{\textbf{WT}} \\\hline
    \textbf{8} & 1.58  & 1.57  & 3.25  & 1.63  & 19.39 & 2.04  & 1.01  & 0.51 \\
    \textbf{16} & 3.26  & 3.17  & 7.00  & 3.50  & 21.11 & 2.29  & 2.09  & 1.08 \\
    \textbf{32} & 6.58  & 6.13  & 15.00 & 7.50  & 23.04 & 2.79  & 4.25  & 2.24 \\
    \textbf{64} & 13.18 & 11.58 & 32.00 & 16.00 & 25.35 & 3.79  & 8.91  & 4.87 \\
    \bottomrule
    \end{tabular}%
  \caption{Numbers of \rwt{s} of different sorting algorithms with various data sizes (bytes).  The input size is $10^7$ and the cache contains 1000 cache-lines.}
  \label{tab:sort-vary-size}%
\bigskip
\end{table}%

\subsection{Conclusions}

Based on our implementations and experiments, the following conclusions and the projection of the techniques for future non-volatile main-memories can be drawn.

\mysubparagraph{Samplesort.}
Samplesort generally requires fewer I/Os than other sorting algorithms.
On existing hardware, since samplesort does not explicitly optimize for L1/L2 cache, its sequential performance is slightly slower than quicksort.
However, in the multicore parallel setting, samplesort is always the fastest due to its efficiency on I/O cost.
We predict that samplesort will play a more significant role with the future hardware even in the sequential setting, because of the lower number of accessing to the large asymmetric memory and the asymmetry of bandwidth and energy on the new memories.

\mysubparagraph{Sorting indirect pointers.}
The advantage of samplesort will be enlarged when sorting via pointers.
This is because samplesort requires fewer rounds to finish, and therefore it uses fewer reads, fewer writes, and fewer comparisons compared to other algorithms.
This also matches the observation on the existing symmetric setting that samplesort is more efficient on wall-clock performance.

\mysubparagraph{BST sort.}
Although BST sort only requiring constant writes on \largemem{} per update, it is still inefficient compared to other approaches when sorting integers or float-point numbers because of its lacking on utilizing the cache-lines.
However, when sorting entries with at least 64 bytes (i.e.\ a cache-line size), the I/O cost of BST sort outperforms the rest three algorithms.

\section{Graph Traversal Algorithms}\label{sec:graph-traversal}

We now provide the full version of the graph-traversal algorithms in this paper.
We discuss two of the most commonly-used graph-traversing algorithms: Breadth-first search (BFS), and Dijkstra's Algorithm.  We show that with appropriate implementations, these algorithms can write much fewer to the main memory, comparing to the classic implementations.

Throughout this section we assume the input graph $G=(V,E)$ contains $n=|V|$ vertices and $m=|E|$ edges.

\subsection{Breadth-First Search}
\label{sec:bfs}

Breadth-first search (BFS) is an algorithm for graph traversing or searching. It starts at the source node, which is an arbitrary node of a graph, and explores the vertices with respect to the distances (the number of hops) from the source node.
BFS is commonly-used in computing single-source shortest paths on unweighted graphs, as a subroutine for graph radii estimation, eccentricity estimation and betweenness centrality, and as a basic building block for other graph algorithms like graph connectivity, reachability, bridges, biconnected components, and strongly connected components.
In this paper we focus on the most basic application: shortest paths on undirected graphs, and the techniques discussed here can apply to many other applications.

\vspace{.5em}
\subsubsection{The classic implementation}

Given a graph $G=(V,E)$ with $n$ vertices and $m$ edges, the classic implementation of BFS keeps a queue with size $n$, and an array of boolean flags with size $n$ indicating that each vertex is visited or not during the search.
This implementation requires at most 2 writes for each vertex: one sequential write for adding it to the queue and one random access for changing the flag of this vertex.
Meanwhile, searches along an edge to an already visited vertex require no writes.
The overall I/O cost of BFS $Q(n,m)=O(\wcost{}n+m)$~\cite{blelloch2016efficient}.

This bound is asymptotic optimal for arbitrary graphs, since the output of BFS, the shortest-path tree, has size $O(n)$.  However, a number applications (e.g.\ s-t shortest-path or connectivity, graph radii estimation or eccentricity estimation) have output size $O(1)$, which allow techniques utilizing the \smallmem{} and reducing the number of writes.

\subsubsection{BFS implementation using \rotating{}}\label{sec:bfs-rotating}

The \rotating{} are used in many algorithms to reduce the space requirement.
For example, the diamond DAG computation in the longest common subsequence (LCS) problem only requires storing two consecutive columns (or rows) at any time during the dynamic programming process, since each DP value only depends on three other nearby vertices in the diamond DAG.
The algorithm maintains two arrays each with the size of either input string: one to hold the results in odd columns (rows) while the other for the even ones.
We call this structure the \emph{\rotating} since the two arrays are rotating and holding computed and uncomputed values.
They only require temporal (cache) space of one dimension although the nodes in the entire DAG have two dimensions.

Here we observe that BFS on undirected graphs can apply a similar approach: instead of keeping a queue and a global array of boolean flags with size $n$, we maintain separate queues, each corresponding to a specific frontier (i.e.\ the set of vertices with the same distance to the source).
This is because, during the BFS on undirected graphs, the outgoing edges from the $i$-th frontier can only reach the vertices in either the $(i-1)$-th frontier or the $(i+1)$-th frontier.
Otherwise, assume an edge reaches a vertex in the $j(<i-1)$-th frontier, then the vertex in $j$-th frontier will visit this vertex in $(j+1)$-th frontier.
As a result, when processing the $i$-th frontier, we only need to keep three frontiers with distance $i-1$, $i$ and $i+1$.

Since each frontier is separately kept and all vertices in one frontier have the same distances, we no longer need to keep the relative orders of the elements within each frontier.
We hence directly use an unordered set to maintain each frontier.
Since only three frontiers are useful during the BFS process, we also only allocate and keep three unordered sets and their roles rotate among the previous, the current and the next frontier.

The 2-level hash table introduced in Section~\ref{sec:hashtable} works perfectly well here since we have no control of the frontier sizes, which can be either greater or smaller than previous ones.
For each frontier, we loop over all vertices in the set, and for each outgoing edge, if the other endpoint is not in either of the three unordered sets then it is added to the next frontier.
Therefore, the operations to the sets include lookups and insertions.
We do not explicitly delete the sets.
Instead, when the searching of one frontier is finished, the new next frontier will reuse this space of the previous frontier.

\subsubsection{Bidirectional search}\label{sec:bfs-bidir}

This previous implementation based on \rotating{} can greatly reduce the number of \wrt{} when the frontier size is much smaller than the number of vertices, like grids, meshes, roadmaps, etc.
For other graphs with larger frontier size and smaller diameter (e.g.\ real-world networks that follow power law), I/O cost is not improved significantly.
However for these graphs, the average distance between every pair of vertices is usually very small, and this is called the ``small-world phenomenon''~\cite{watts1998collective}.

To utilize this property, we employ bidirectional breadth-first search on computing the s-t shortest path.
Assuming the distance between this pair of query vertices is $d$, then the search from each direction will only visit the vertices within distance $\lceil d/2\rceil$.
On these power-law graphs, the frontier size grows exponentially on the first several steps until most of the vertices are reached.
The number of vertices on the top-half levels in the shortest-path tree is therefore far less than $n$.
This difference however, is negligible on graphs with bounded frontier size, since the distance between a random pair of vertices is large expectedly, and the bidirectional search eventually reaches about the same amount of vertices unless $d$ is much smaller than the graph diameter.

To implement bidirectional search efficiently, we also use three \rotating{} for the search process.
The extra detail here is that, the source, destination, and all intermediate vertices are put together in the same sets, and use one bit (the sign bit) to identify if the vertex is by the search from the source or from the destination.
We only keep three instead of six sets, which reduces the number of read transfers approximately by a half.

\subsubsection{Experiment}
\label{sec:graph}

In the experiment, we examine whether and how the new algorithms we just discussed
improve the I/O cost.
We implement four versions of breadth-first search: classic and bidirectional search with and without using \rotating{}.
The experiment is run on 8 graph instances with various cache size.

\myparagraph{Graph instances.}\label{sec:graph}
We use graphs of various types from the SNAP datasets~\cite{leskovec2014snap} as the input of the algorithms.
The graphs include the road networks in Pennsylvania and Texas (real-world planar graphs),
the web graphs of the University of Notre Dame and Stanford University, the DBLP collaboration network, and the Youtube online social network
(4 real-world networks).  In the case of web graphs, each edge represents a
hyperlink between two web pages.
We also use synthetic graphs of 2D and 3D grids.
For each of the graphs used, the numbers of vertices and edges are given in the table below.
If a graph does not come equipped with weights, we assign to every edge a random integer between $1$ and $10,000$.
The graphs we used are relatively small due to the overhead in our software and hardware simulator, but we adjust the cache size accordingly.

\begin{table}[!h]
\def\arraystretch{1.1}
\centering
\small
\begin{tabular}{c|l|rr}
\toprule
  Type & \multicolumn{1}{c|}{Graph} & \# Vertices & \# Edges \\
\hline
  & Grid: 2D & 1M & 3.96M \\
Sparse  & Grid: 3D & 1M & 5.94M \\
Graphs  & Roadmap: Pennsylvania & 1.09M & 3.08M \\
  & Roadmap: Texas & 1.39M & 3.84M \\ 
  \hline
  & Webgraph: Notre Dame & 325K & 2.20M\\
Social  & Webgraph: Stanford & 281K & 3.98M\\
Networks  & DBLP collab.\ network & 317K & 2.10M\\
  & Youtube network & 1.13M & 5.98M\\
\bottomrule
\end{tabular}
\end{table}

\begin{figure*}[!t]
\begin{center}
    \begin{minipage}[t]{0.48\textwidth}
        \includegraphics[width=\textwidth]{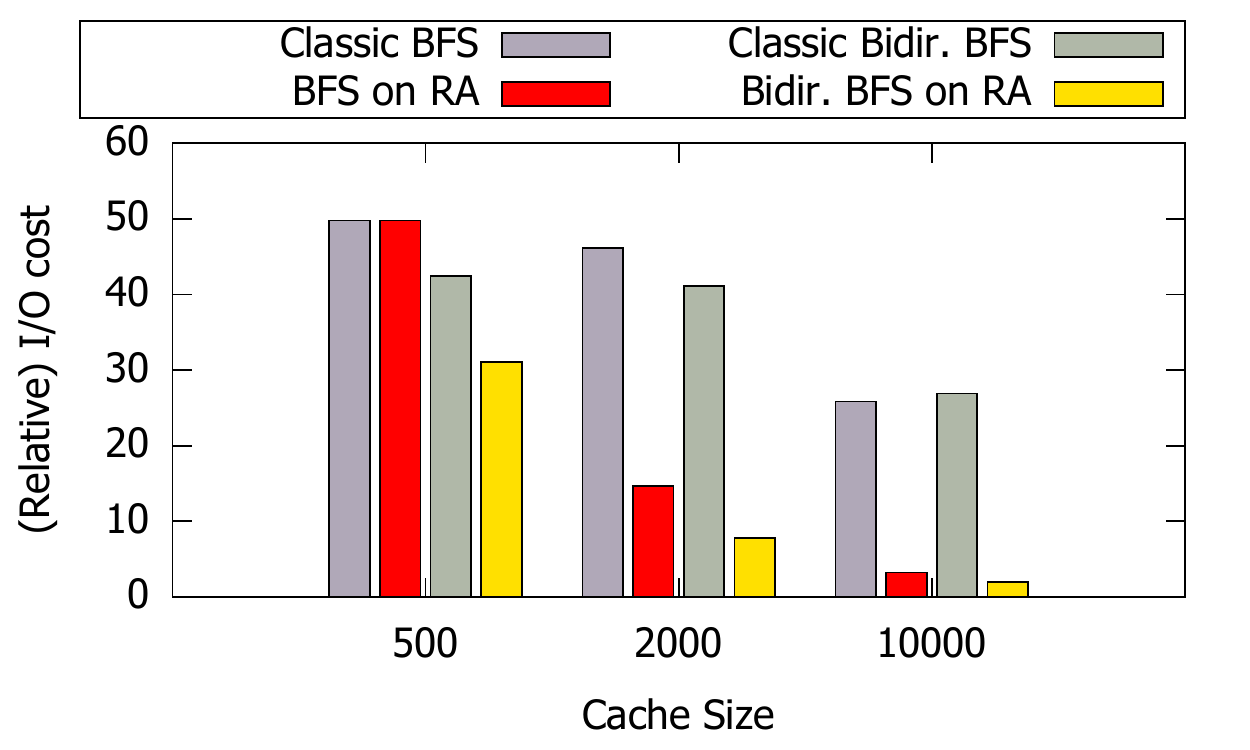} \centering{\small (a) $\wcost=10$, sparse graphs}
    \end{minipage}
    \begin{minipage}[t]{0.48\textwidth}
        \includegraphics[width=\textwidth]{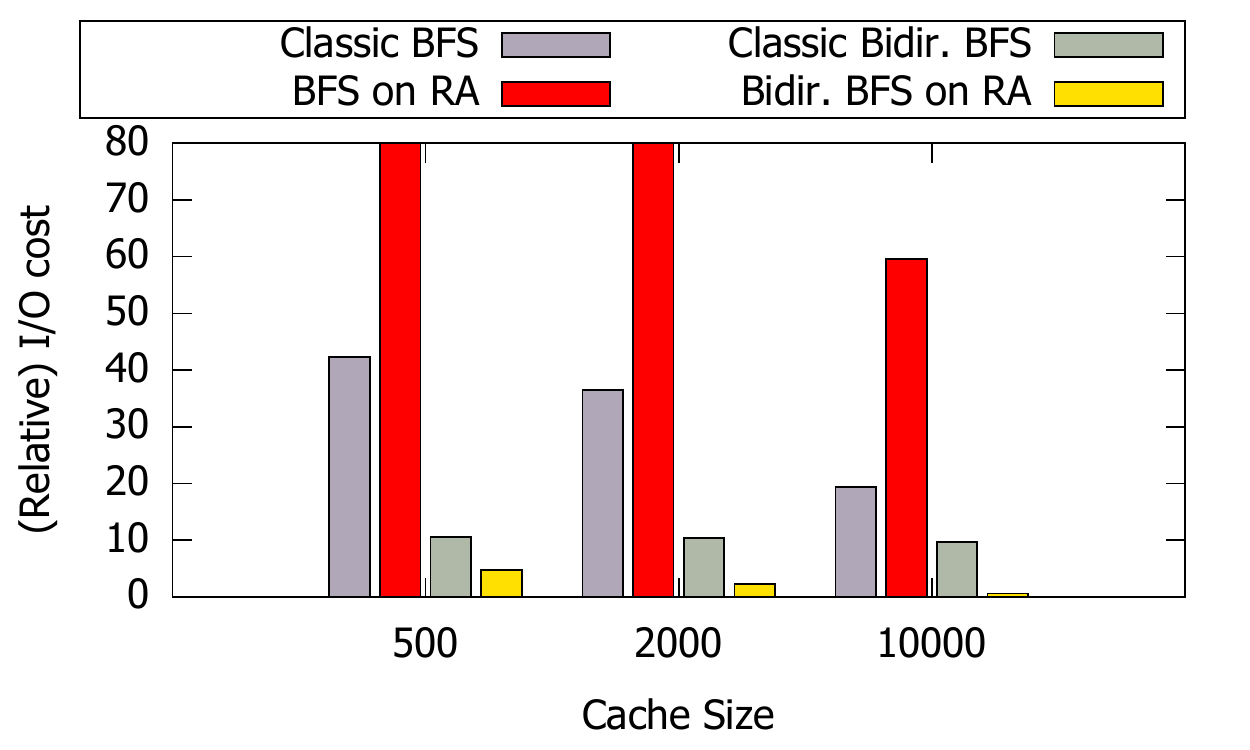} \centering{\small (b) $\wcost=10$, social networks}
    \bigskip
    \end{minipage}
    \begin{minipage}[t]{0.48\textwidth}
        \includegraphics[width=\textwidth]{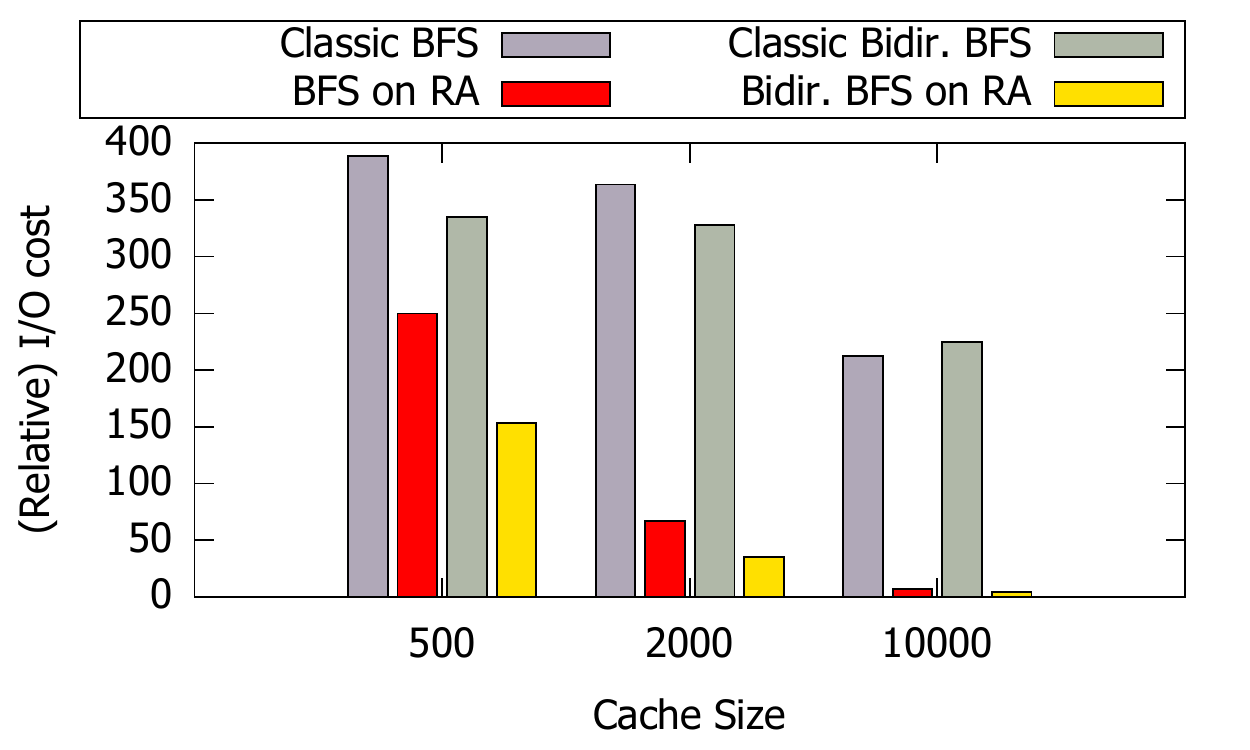} \centering{\small (c) $\wcost=100$, sparse graphs}
    \end{minipage}
    \begin{minipage}[t]{0.48\textwidth}
        \includegraphics[width=\textwidth]{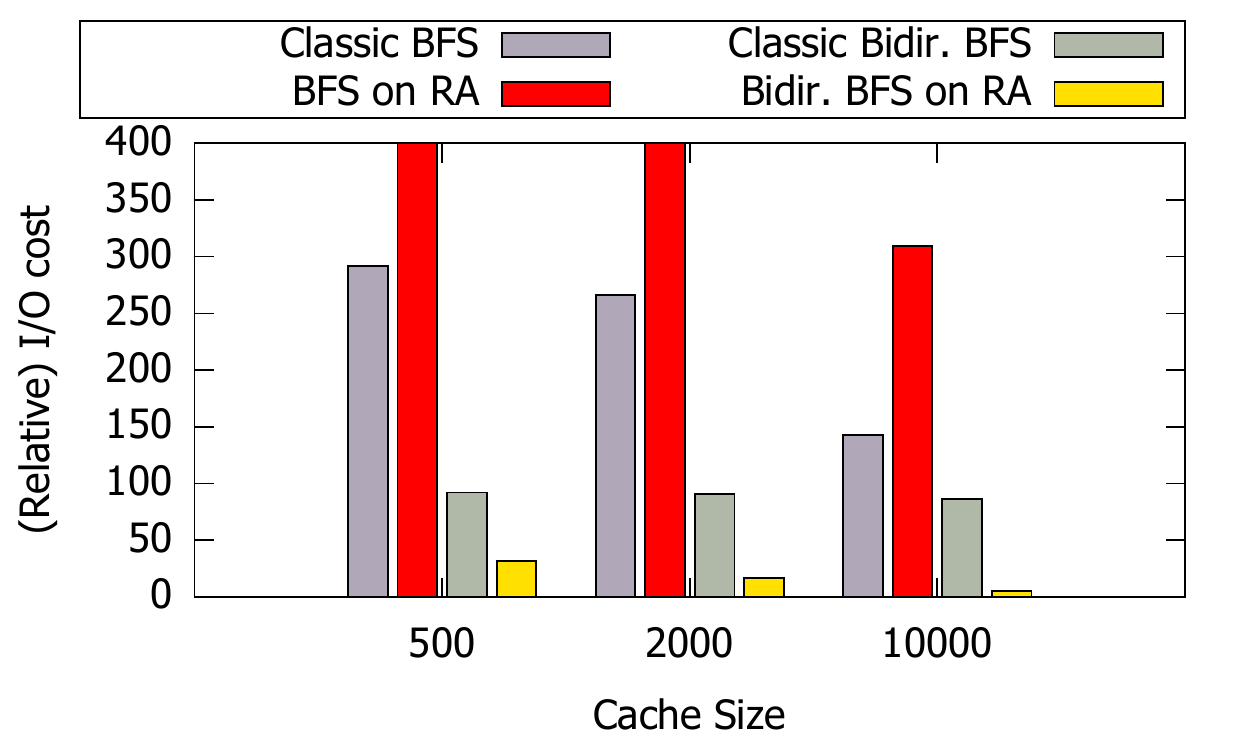} \centering{\small (d) $\wcost=100$, social networks}
    \end{minipage}
    \captionof{figure}{\label{fig:bfs} The trends of the I/O costs of four different implementations of BFS.
    The new implementations shown in this paper are the BFS and bidirectional BFS based on \rotating{} (red and orange bars).
    Graphs used in the experiment are shown in Section~\ref{sec:graph} and categorized into sparse (almost planar) graphs and social networks.
    We show the relative I/O cost based on varied cache sizes, and each number is geometric mean of the four graphs in that category (the exact numbers are given in Table~\ref{tbl:bfs-iocost}).
    We can see the consistent advantages of the new BFS implementation on sparse graphs, and the improvement of the new bidirectional version in all cases.
    Notice that in (b) and (d) some values exceed the ranges of vertical axis.
    }
\end{center}
\end{figure*}

\mysubparagraph{Overall performance.}
The numbers of \rwt{s} on 8 graph instances with various cache size are given in Table~\ref{tbl:bfs-raw}.
Their weighted I/O costs are given in Table~\ref{tbl:bfs-iocost} considering two typical values of $\wcost$, and visualized (merged into two categories) in Figure~\ref{fig:bfs}.

To better understand the performance of different implementations, we also count various statistics of the searching including the depths and frontier sizes, and the results are generalized in Table~\ref{tab:BFS-stats}.

\begin{table*}[!t]
\centering
\footnotesize
\def\arraystretch{1.2}
    \begin{tabular}{l|rrrrrr|rrrrrr}
    \toprule
    \multicolumn{1}{c|}{\textbf{Algorithm}} & \multicolumn{6}{c|}{\textbf{Classic BFS}}      & \multicolumn{6}{c}{\textbf{BFS based on RA}} \\
    \cline{1-13}
    \multicolumn{1}{c|}{\textbf{Cache Size}} & \multicolumn{2}{c}{\textbf{500}} & \multicolumn{2}{c}{\textbf{2000}} & \multicolumn{2}{c|}{\textbf{10000}} & \multicolumn{2}{c}{\textbf{500}} & \multicolumn{2}{c}{\textbf{2000}} & \multicolumn{2}{c}{\textbf{10000}} \\
    \cline{1-13}
    \textbf{} & \multicolumn{1}{c}{\textbf{RT}} & \multicolumn{1}{c}{\textbf{WT}} & \multicolumn{1}{c}{\textbf{RT}} & \multicolumn{1}{c}{\textbf{WT}} & \multicolumn{1}{c}{\textbf{RT}} & \multicolumn{1}{c|}{\textbf{WT}} & \multicolumn{1}{c}{\textbf{RT}} & \multicolumn{1}{c}{\textbf{WT}} & \multicolumn{1}{c}{\textbf{RT}} & \multicolumn{1}{c}{\textbf{WT}} & \multicolumn{1}{c}{\textbf{RT}} & \multicolumn{1}{c}{\textbf{WT}} \\
    \hline

    \textbf{2D Grid} & 18.47 & 6.17  & 17.60 & 6.01  & 10.88 & 4.59  & 8.75  & 0.84  & 6.52  & 0.16  & 6.20  & 0.01 \\
    \textbf{3D Grid} & 19.38 & 5.45  & 16.42 & 5.31  & 13.51 & 4.51  & 88.24 & 4.26  & 31.03 & 2.01  & 6.39  & 0.38 \\
    \textbf{PA Roadmap} & 8.73  & 2.78  & 8.07  & 2.54  & 2.43  & 1.05  & 29.61 & 2.95  & 5.37  & 0.65  & 1.26  & 0.02 \\
    \textbf{TX Roadmap} & 6.79  & 2.16  & 6.01  & 1.91  & 1.95  & 0.85  & 23.45 & 2.29  & 4.32  & 0.50  & 1.00  & 0.02 \\
    \textbf{Stan Webgraph} & 39.75 & 3.97  & 27.11 & 3.75  & 9.63  & 1.48  & 232.09 & 5.99  & 150.70 & 4.74  & 35.19 & 2.10 \\
    \textbf{NDU Webgraph} & 3.47  & 1.04  & 3.02  & 0.97  & 2.52  & 0.77  & 84.67 & 4.80  & 58.91 & 3.91  & 18.33 & 1.73 \\
    \textbf{DBLP Network} & 19.70 & 5.75  & 15.43 & 5.07  & 6.85  & 1.68  & 140.01 & 7.34  & 104.12 & 6.27  & 37.67 & 3.20 \\
    \textbf{Youtube Network} & 13.23 & 2.47  & 9.64  & 2.29  & 5.63  & 1.84  & 91.70 & 6.77  & 71.98 & 6.39  & 39.95 & 5.08 \\

    \midrule
    \multicolumn{1}{c|}{\textbf{Algorithm}} & \multicolumn{6}{c|}{\textbf{Classic Bidirectional BFS}}      & \multicolumn{6}{c}{\textbf{Bidirectional BFS based on RA}} \\
    \cline{1-13}
    \multicolumn{1}{c|}{\textbf{Cache Size}} & \multicolumn{2}{c}{\textbf{500}} & \multicolumn{2}{c}{\textbf{2000}} & \multicolumn{2}{c|}{\textbf{10000}} & \multicolumn{2}{c}{\textbf{500}} & \multicolumn{2}{c}{\textbf{2000}} & \multicolumn{2}{c}{\textbf{10000}} \\
    \cline{1-13}
    \textbf{} & \multicolumn{1}{c}{\textbf{RT}} & \multicolumn{1}{c}{\textbf{WT}} & \multicolumn{1}{c}{\textbf{RT}} & \multicolumn{1}{c}{\textbf{WT}} & \multicolumn{1}{c}{\textbf{RT}} & \multicolumn{1}{c|}{\textbf{WT}} & \multicolumn{1}{c}{\textbf{RT}} & \multicolumn{1}{c}{\textbf{WT}} & \multicolumn{1}{c}{\textbf{RT}} & \multicolumn{1}{c}{\textbf{WT}} & \multicolumn{1}{c}{\textbf{RT}} & \multicolumn{1}{c}{\textbf{WT}} \\
    \hline

    \textbf{2D Grid} & 18.14 & 4.97  & 17.61 & 4.94  & 9.39  & 4.61  & 7.60  & 0.55  & 4.94  & 0.13  & 4.69  & 0.01 \\
    \textbf{3D Grid} & 11.66 & 3.08  & 10.52 & 3.06  & 9.47  & 2.98  & 38.24 & 1.97  & 8.80  & 0.69  & 2.90  & 0.09 \\
    \textbf{PA Roadmap} & 7.87  & 3.19  & 7.58  & 3.09  & 3.37  & 1.60  & 27.00 & 2.62  & 4.25  & 0.43  & 1.06  & 0.02 \\
    \textbf{TX Roadmap} & 5.31  & 2.30  & 5.03  & 2.19  & 1.79  & 1.06  & 11.12 & 1.18  & 2.30  & 0.21  & 0.60  & 0.01 \\
    \textbf{Stan Webgraph} & 4.70  & 1.38  & 4.24  & 1.37  & 2.74  & 1.29  & 7.32  & 0.84  & 3.00  & 0.47  & 0.63  & 0.13 \\
    \textbf{NDU Webgraph} & 0.80  & 0.70  & 0.77  & 0.70  & 0.74  & 0.68  & 2.63  & 0.43  & 1.12  & 0.28  & 0.16  & 0.06 \\
    \textbf{DBLP network} & 1.40  & 1.01  & 1.29  & 0.99  & 1.05  & 0.89  & 1.33  & 0.25  & 0.37  & 0.12  & 0.12  & 0.04 \\
    \textbf{Youtube network} & 0.74  & 0.68  & 0.71  & 0.68  & 0.69  & 0.67  & 0.25  & 0.09  & 0.08  & 0.04  & 0.02  & 0.01 \\

    \bottomrule
\end{tabular}
\caption{\label{tbl:bfs-raw}
Numbers of read and write transfers of BFS implementations on different cache sizes.  Numbers are r/w transfers per vertex per \textbf{10} queries.  We pick the number 10 to fit the numbers in one table.}
\bigskip
\end{table*}
\begin{table*}[!t]
\centering
\footnotesize
\def\arraystretch{1.25}
    \begin{tabular}{l|r@{  }@{  }r@{  }@{  }r|r@{  }@{  }r@{  }@{  }r||r@{  }@{  }r@{  }@{  }r|r@{  }@{  }r@{  }@{  }r}
    \toprule
    \multicolumn{13}{c}{$\wcost{}=10$} \\
    \hline
    \textbf{Algorithm} & \multicolumn{3}{c|}{\textbf{Classic BFS}} & \multicolumn{3}{c||}{\textbf{BFS based on RA}} & \multicolumn{3}{c|}{\textbf{Classic Bidir.\ BFS}} & \multicolumn{3}{@{ }c@{}}{\textbf{Bidir.\ BFS based on RA}} \\
    \hline
    \textbf{Cache Size} & \multicolumn{1}{c}{\textbf{500}} & \multicolumn{1}{c}{\textbf{2000}} & \multicolumn{1}{@{}c|}{\textbf{10000}} & \multicolumn{1}{c}{\textbf{500}} & \multicolumn{1}{c}{\textbf{2000}} & \multicolumn{1}{@{}c||}{\textbf{10000}} & \multicolumn{1}{c}{\textbf{500}} & \multicolumn{1}{c}{\textbf{2000}} & \multicolumn{1}{@{}c|}{\textbf{10000}} & \multicolumn{1}{c}{\textbf{500}} & \multicolumn{1}{c}{\textbf{2000}} & \multicolumn{1}{@{}c@{}}{\textbf{10000}} \\
    \hline
    \textbf{2D Grid} & 80.18 & 77.72 & 56.74 & 17.13 & 8.15  & 6.25  & 67.80 & 67.04 & 55.53 & 13.14 & 6.27  & 4.76 \\
    \textbf{3D Grid} & \textit{73.85} & 69.49 & 58.60 & 130.86 & 51.16 & 10.18 & \textit{42.44} & 41.16 & 39.29 & 57.98 & 15.65 & 3.82 \\
    \textbf{PA Roadmap} & \textit{36.50} & 33.42 & 12.89 & 59.10 & 11.84 & 1.44  & \textit{39.76} & 38.46 & 19.39 & 53.24 & 8.52  & 1.24 \\
    \textbf{TX Roadmap} & \textit{28.42} & 25.14 & 10.44 & 46.36 & 9.36  & 1.18  & 28.27 & 26.95 & 12.37 & 22.89 & 4.37  & 0.69 \\
    \textbf{Stan Webgraph} & \textit{79.41} & \textit{64.64} & \textit{24.39} & 291.99 & 198.08 & 56.18 & 18.48 & 17.93 & 15.62 & 15.75 & 7.71  & 1.93 \\
    \textbf{NDU Webgraph} & \textit{13.84} & \textit{12.70} & \textit{10.20} & 132.71 & 98.04 & 35.58 & 7.80  & 7.73  & 7.54  & 6.90  & 3.90  & 0.75 \\
    \textbf{DBLP network} & \textit{77.19} & \textit{66.16} & \textit{23.67} & 213.43 & 166.84 & 69.63 & 11.53 & 11.22 & 9.91  & 3.84  & 1.61  & 0.56 \\
    \textbf{Youtube network} & \textit{37.92} & \textit{32.59} & \textit{24.06} & 159.43 & 135.92 & 90.72 & 7.58  & 7.50  & 7.38  & 1.17  & 0.51  & 0.17 \\
    \midrule
        \multicolumn{13}{c}{$\wcost{}=100$} \\
    \hline
    \textbf{Algorithm} & \multicolumn{3}{c|}{\textbf{Classic BFS}} & \multicolumn{3}{c||}{\textbf{BFS based on RA}} & \multicolumn{3}{c|}{\textbf{Classic Bidir.\ BFS}} & \multicolumn{3}{@{ }c@{}}{\textbf{Bidir.\ BFS based on RA}} \\
    \hline
    \textbf{Cache Size} & \multicolumn{1}{c}{\textbf{500}} & \multicolumn{1}{c}{\textbf{2000}} & \multicolumn{1}{@{}c|}{\textbf{10000}} & \multicolumn{1}{c}{\textbf{500}} & \multicolumn{1}{c}{\textbf{2000}} & \multicolumn{1}{@{}c||}{\textbf{10000}} & \multicolumn{1}{c}{\textbf{500}} & \multicolumn{1}{c}{\textbf{2000}} & \multicolumn{1}{@{}c|}{\textbf{10000}} & \multicolumn{1}{c}{\textbf{500}} & \multicolumn{1}{c}{\textbf{2000}} & \multicolumn{1}{@{}c@{}}{\textbf{10000}} \\
    \hline
    \textbf{2D Grid} & 635.5 & 618.7 & 469.4 & 92.5 & 22.7 & 6.7  & 514.7 & 511.8 & 470.7 & 62.9 & 18.2 & 5.3 \\
    \textbf{3D Grid} & 564.0 & 547.1 & 464.3 & 514.4 & 232.3 & 44.2 & 319.4 & 316.9 & 307.6 & 235.6 & 77.4 & 12.0 \\
    \textbf{PA Roadmap} & \textit{286.4} & 261.5 & 107.0 & 324.4 & 70.0 & 3.0  & 326.7 & 316.4 & 163.6 & 289.4 & 47.0 & 2.8 \\
    \textbf{TX Roadmap} & \textit{223.1} & 197.3 &  86.8 & 252.5 & 54.7 & 2.8  & 234.8 & 224.2 & 107.6 & 128.8 & 23.0 & 1.5 \\
    \textbf{Stan Webgraph} & \textit{436.3} & \textit{402.4} & 157.1 & 831.1 & 624.5 & 245.1  & 142.5 & 141.2 & 131.6 & 91.6 & 50.1 & 13.6 \\
    \textbf{NDU Webgraph} & \textit{107.1}  &  \textit{99.8} & 79.2 & 565.0 & 450.2 & 190.8   &  70.8 &  70.3 & 68.8 & 45.3 & 28.9 & 6.0 \\
    \textbf{DBLP network} & \textit{594.6}  & \textit{522.6} & 175.1 & 874.2 & 731.3 & 357.3  & 102.6 & 100.5 & 89.7 & 26.4 & 12.8 & 4.5 \\
    \textbf{Youtube network} & \textit{260.2} & \textit{239.1} & 190.0 & 769.0 & 711.4 & 547.6 & 69.1 &  68.5 & 67.6 &  9.4  & 4.3  & 1.5 \\
    \bottomrule
\end{tabular}
\caption{\label{tbl:bfs-iocost}
I/O costs of BFS implementations on different cache sizes.  Numbers of r/w transfers are from Table~\ref{tbl:bfs-raw}.  Italic-font number indicates that the classic implementation has a lower I/O cost in that setting.}
\bigskip
\end{table*}

\begin{table*}[!ht]
\centering
\small
\def\arraystretch{1.25}
    \begin{tabular}{l|rrr|rrr}
    \toprule
          & \multicolumn{3}{c|}{\textbf{Unidirectional search}} & \multicolumn{3}{c}{\textbf{Bidirectional search}} \\
    \cline{2-7}
          & \multicolumn{1}{p{.8cm}<{\centering}}{\textbf{depth}} & \multicolumn{1}{p{1.2cm}<{\centering}}{\textbf{maximum frontier size}} & \multicolumn{1}{p{1.2cm}<{\centering}|}{\textbf{average frontier size}} & \multicolumn{1}{p{1.2cm}<{\centering}}{\textbf{depth}} & \multicolumn{1}{p{1.2cm}<{\centering}}{\textbf{maximum frontier size}} & \multicolumn{1}{p{1.2cm}<{\centering}}{\textbf{average frontier size}} \\
    \hline
    \textbf{2D Grid} & 1571.4 & 1460  & 636.4 & 370.2 & 2091  & 1102.8 \\
    \textbf{3D Grid} & 229.9 & 13277 & 4349.7 & 49.3  & 15058 & 4628.9 \\
    \textbf{PA Roadmap} & 574.5 & 5032  & 1893.1 & 163.9 & 7204  & 2718.3 \\
    \textbf{TX Roadmap} & 748.6 & 6058  & 1804.9 & 174.7 & 5752  & 2102.1 \\
    \textbf{Stan Webgraph} & 107.1 & 87614 & 2383.4 & 3.5   & 50094 & 3782.8 \\
    \textbf{NDU Webgraph} & 29.2  & 124519 & 11155.1 & 3.8   & 51562 & 2053.5 \\
    \textbf{DBLP Network} & 16.1  & 129283 & 19694.4 & 3.9   & 36373 & 1420.1 \\
    \textbf{Youtube Network} & 15.8  & 607534 & 71828.5 & 2.7   & 1841  & 107.4 \\
    \bottomrule
    \end{tabular}%
  \caption{The depths (number of levels in the shortest-path tree) and frontier sizes (number of vertices) during the search processes.  Note that the numbers are averaged from multiple searches.}
  \label{tab:BFS-stats}%
\bigskip  
\end{table*}%

\myparagraph{Unidirectional search.}
Indicated in Table~\ref{tbl:bfs-raw},
classic BFS requires no more than one sequential write (pushed into the queue) and one random write (marking the distance) per vertex.
The random write can be avoided if the associated cache line is cached, so better locality of vertex indices of some graphs (roadmaps and NDU webgraph) and larger cache size reduce writes per vertex.
The algorithm also reads the edges of each vertex and checks whether the other endpoint is visited or not, which is also affected by the edges per vertex and the locality of vertex indices.

For the implementation using \rotating{}, the key factor is whether the frontier fits into the cache.
Shown in Table~\ref{tab:BFS-stats}, the sparse graphs (grids and roadmaps) have smaller frontier sizes, so as long as the cache can hold each of them, the writes are largely minimized.
This is also true for reads since checking is always in the hash table.
However, once the frontier is larger than the cache size, then each insertion to the hash table now becomes a random write and can hardly be cached because of the hash function.
The reads are increased even more, since checking whether a vertex is visited can lead to at most 6 cache misses: three \rotating{} each with 2 levels.
Summarized in Figure~\ref{fig:bfs}, the I/O cost is largely improved using the \rotating{} when frontiers fit into the cache, but deteriorates when not.

\myparagraph{Bidirectional search.}
The I/O performance of classic bidirectional BFS is similar to the classic unidirectional BFS since they essentially search in a similar pattern.
The bidirectional search requires fewer reads and writes since it strictly searches fewer vertices, especially in social networks since these graphs have smaller diameters and the two searches usually meet earlier before the majority of the vertices are visited.

This special property of social network largely helps our implementation using \rotating{}.
Shown in Table~\ref{tab:BFS-stats}, the two bidirectional searches use 2.7-3.9 rounds on average to meet, which preserves the frontier in the cache during the search even for very small cache sizes.
As a result, the bidirectional BFS using \rotating{} has a constantly good performance on all combinations of graphs and cache sizes, which is shown in Table~\ref{tbl:bfs-iocost} and Figure~\ref{fig:bfs}.

\subsubsection{Conclusions}

We discuss how to efficiently implement BFS in the asymmetric setting and experiment the I/O performance for four implementations on a variety of graphs.
We show that if the query is s-t (pairwise) distance, our bidirectional BFS using \rotating{} shows an overwhelming advantage in all cases we tested.
If all distances to the source are required, the unidirectional BFS using \rotating{} has a better performance if the cache can hold each frontier. 

\subsection{Dijkstra's Algorithm}
\label{sec:dijk}

Dijkstra's Algorithm~\cite{dijkstra1959} is a well-known algorithm to compute single-source shortest paths on a non-negative weighted graph $G=(V,E)$.
Due to the rapid growth of the data size, real-world graphs nowadays can easily go beyond the size of the order of gigabytes, and they need to be stored on the large non-volatile memory.
Running the classic implementation of Dijkstra's algorithm can be costly in this setting.
We show that with an appropriate implementation, the algorithm can write much fewer to the non-volatile memory, which further leads to much lower I/O cost.

Throughout this section we assume the input graph $G=(V,E)$ contains $n=|V|$ vertices and $m=|E|$ edges.

Dijkstra's algorithm maintains a set of visited vertices associated with their shortest distances to the source (denoted by $d_v$ for $v\in V$), and the unvisited neighbors of these vertices form the ``frontier'' ($d_v$ for unvisited vertex $v$ is $+\infty$).
Each vertex $u$ in the frontier stores a tentative distance to be $\min_{v\in N(u)}\{d_v+e_{u,v}\}$ where $N(u)$ is the incoming neighbor set of vertex $u$.
Initially the visited set contains only one vertex: the source node.
The invariant of this algorithm is that, the minimum tentative distance in the frontier set is indeed the shortest distance of this vertex, and thus Dijkstra's algorithm iteratively move this vertex from the frontier to the visited set and update the frontier accordingly.
The correctness can easily be proven by induction on the number of visited nodes.

Due to the widespread use of Dijkstra's algorithm, there are plenty of works on the efficient implementations of Dijkstra's algorithm and we refer the readers to some recent works~\cite{MeyerSanders2003,blelloch2016parallel} for summaries of the work bounds of different approaches.
Different implementations have different costs on the two operations in Dijkstra's algorithm: \mf{Extract-Min} that finds and removes the vertex with minimum distance in the frontier set, and \mf{Decrease-Key} that updates the tentative distance of the other endpoint of an edge of this vertex removed from the frontier in this iteration.
The data type that supports the queries is abstracted as a priority queue.
Specifically when the priority queue is implemented using Fibonacci Heap~\cite{fredman1987fibonacci} to maintain the frontier set, the overall time complexity is $O(m+n\log n)$.

\subsubsection{Classic implementation using a binary heap}

Although there are many advanced implementations of the priority queue with lower time complexities, the constant hidden in the asymptotic bound is large.
In practice the classic implementation using a binary heap works reasonably well on general sparse real-world graphs, and its wall-clock performance is competitive or even better on modern computer architecture.
We hence implement it as a baseline and measure the number of read and write transfers of this algorithm as a comparison to our write-efficient version.

For each vertex, we maintain the shortest distance in a global array with size $n$.
For practical purpose, instead of initializing the priority queue of size $n$ with infinite distances, we insert a vertex when it is first added to the frontier (so the priority queue needs to support \mf{Insert}, which most implementations do).
The binary heap only stores the indices of the vertices which optimizes the memory footprint and number of \wrt{s}.
To perform \mf{Decrease-Key} in a binary heap efficiently, we keep another global array, an auxiliary structure that maps each vertex to its position in the heap, and is maintained up-to-date as the priority queue changes.
This implementation has worst-case time complexity to be $O(m\log n)$.  Since this implementation does not take any special optimization on caching, the I/O cost is therefore $O(\wcost{}m\log (nB/M))$, assuming the cache always keeps the first $M/B$ vertices in the priority queue.

\subsubsection{Phased Dijkstra}

Phased Dijkstra~\cite{blelloch2016efficient} is a specific implementation of Dijkstra's algorithm that the goal is to  fully maintain the priority queue in \smallmem{}.   It only requires $O(n)$ writes to large-memory, the shortest distances to all vertices.
The idea is to partition the computation into
phases such that for a parameter $M'$
each phase keeps a priority queue of size at most $(1+\epsilon)M'$
and visits at least
$M'$ vertices. By selecting $M' = M/c$ for an appropriate constant $c$, the priority queue fits in \smallmem{},
and the only writes to \largemem{} are the final distances.

\begin{algorithm}[ht]
\small
\caption{Phased Dijkstra}
\label{algo:wo-dijkstra}

\fontsize{10pt}{10pt}\selectfont
    \KwIn{A connected weighted graph $G=(V,E)$ and a source vertex $s$}
    \KwOut{The shortest distances $\delta=\{\delta_1,\ldots, \delta_n\}$ from source $s$}
    \vspace{0.5em}

{Priority Queue $P\leftarrow \varnothing$}\\
{Mark vertex $s$ as visited and set $\delta(s)\leftarrow 0$}\\
\While {\upshape there exists unvisited vertices} {
\If {$P = \varnothing$} {
    {Scan over all edges in $E$ and store at most $\memsize'$ closest unvisited vertices in $P$} \label{line1}\\
    \If {$|P|=\memsize'$} {
        {Set $d_{\smb{max}}$ as the distance to the farthest vertex in $P$}
    }
        \Else {$d_{\smb{max}}\leftarrow +\infty$\label{line2}}
    }
{$u=P.\mf{Extract-Min}()$ \label{line3}}\\
{Set $\delta_u$ as the distance from $s$ to $u$, and mark $u$ as visited \label{line5}}\\
\ForEach {\upshape $(u,v,\mb{dis}_{u,v})\in E$} {
    \If {\upshape $\delta_u+\mb{dis}_{u,v}<d_{\smb{max}}$} {
        \If {$v\in P$\label{check}} {
            {$P.\mf{Decrease-Key}(v,\delta_u+\mb{dis}_{u,v})$}
        }
        \Else {
            {$P.\mf{insert}(v,\delta_u+\mb{dis}_{u,v})$}\\
            \If {$|P| = (1+\epsilon)\memsize'$} {
                {Remove $\epsilon\memsize'$ vertices with larger distances in $P$}\\
                {Set $d_{\smb{max}}$ as the farthest distance in $P$\label{line4}}
            }
        }
    }
}
}
\Return $\delta(\cdot)$
\end{algorithm}

The pseudocode of the algorithm is provided in
Algorithm~\ref{algo:wo-dijkstra}.
Technically the priority queue $P$ can be implemented using an arbitrary heap since it is fully in the \smallmem{} and will not affect the I/O cost.
In our experiment we implement it using a binary heap.

In phased Dijkstra, each phase starts and ends with an empty priority queue $P$.
The priority queue is kept
small by discarding the $\epsilon\memsize'$ largest elements (vertex distances)
whenever $|P| = (1+\epsilon)\memsize'$.  To achieve this, $P$ is flattened into an
array, the $\memsize'$-th smallest element $d_\smb{max}$ is found by
selection, and the priority queue is reconstructed from the elements
no greater than $d_\smb{max}$, all taking linear time.  After such a truncation, all further
insertions in a given phase are not added to $P$ if they have a value
greater than $d_\smb{max}$.

The processing of each phase in phased Dijkstra consists of two parts.
The first part (line~\ref{line1}--\ref{line2}) of each phase loops over all
edges in the graph and relaxes any that go from a visited to an
unvisited vertex (possibly inserting or decreasing a key in $P$).  The
second part (repeatedly loop over line~\ref{line3}--\ref{line4}) then runs standard Dijkstra's algorithm repeatedly
visiting the vertex with minimum distance and relaxing its neighbors
until $P$ is empty.  Similar to other implementations, to implement relax, the algorithm needs to know
whether a vertex is already in $P$, and if so where in $P$ so that it
can do a decrease-key on it.  However in phased Dijkstra it is too costly to store this
information using a global array.  Instead, we use an unordered map introduced in Section~\ref{sec:hashtable} for this mapping.  Theoretically the hash table can be set with fixed size, but in practice we use a 2-level hash table since it leads to better performance when the frontier size is consistently small, and equal performance otherwise.  This map is referred as \defn{vertex map} later.

The I/O cost of phased Dijkstra is $\displaystyle Q(n,m)=O\left(n\left({m\over\memsize}+\wcost{}\right)\right)$.  More details on the correctness and complexity analysis can be found in~\cite{blelloch2016efficient}.

\begin{figure*}[!p]
\centering
\footnotesize
\def\arraystretch{1.2}
    \begin{tabular}{l|r@{  }@{ }r|r@{  }@{ }r|r@{  }@{ }r|r@{  }r@{ }@{ }r|r@{  }@{  }r@{ }r|r@{  }@{  }r@{ }r} \toprule
          & \multicolumn{6}{c|}{\textbf{Classic Dijkstra using binary heap}} & \multicolumn{9}{c}{\textbf{Phased Dijkstra}} \\
\cline{1-16}
    \multicolumn{1}{c|}{\textbf{Cache Size}} & \multicolumn{2}{c|}{\textbf{2k}} & \multicolumn{2}{c|}{\textbf{10k}} & \multicolumn{2}{c|}{\textbf{50k}} & \multicolumn{3}{c|}{\textbf{2k}} & \multicolumn{3}{c|}{\textbf{10k}} & \multicolumn{3}{c}{\textbf{50k}} \\
\hline
    \textbf{Graph Instance} & \multicolumn{1}{c}{\textbf{RT}} & \multicolumn{1}{c|}{\textbf{WT}} & \multicolumn{1}{c}{\textbf{RT}} & \multicolumn{1}{c|}{\textbf{WT}} & \multicolumn{1}{c}{\textbf{RT}} & \multicolumn{1}{c|}{\textbf{WT}} & \multicolumn{1}{c}{\textbf{RT}} & \multicolumn{1}{@{}c@{}}{\textbf{WT}} &  & \multicolumn{1}{c}{\textbf{RT}} & \multicolumn{1}{@{}c@{}}{\textbf{WT}} & & \multicolumn{1}{c}{\textbf{RT}} & \multicolumn{1}{@{}c@{}}{\textbf{WT}} \\
    \hline
    \textbf{2D Grid} & 10.37 & 4.87  & 4.61  & 2.34  & 3.53  & 1.89  & 5.93  & 1.12  & (1)   & 3.62  & 1.11  & (1)   & 2.74  & 1.04  & (1) \\
    \textbf{3D Grid} & 25.87 & 14.53 & 17.44 & 8.54  & 8.20  & 3.35  & 126.56 & 1.12  & (383) & 28.16 & 1.12  & (44)  & 5.90  & 1.07  & (1) \\
    \textbf{PA Roadmap} & 7.82  & 4.14  & 2.16  & 0.81  & 1.12  & 0.40  & 30.40 & 0.99  & (202) & 1.89  & 0.52  & (1)   & 1.04  & 0.28  & (1) \\
    \textbf{TX Roadmap} & 7.69  & 4.10  & 2.16  & 0.82  & 1.09  & 0.40  & 37.93 & 0.97  & (262) & 1.88  & 0.53  & (1)   & 1.01  & 0.27  & (1) \\
    \textbf{Stan Webgraph} & 37.31 & 18.13 & 26.73 & 12.06 & 10.60 & 3.91  & 404.61 & 1.03  & (100) & 81.02 & 1.02  & (16)  & 13.83 & 0.88  & (2) \\
    \textbf{NDU Webgraph} & 24.97 & 15.34 & 15.14 & 8.21  & 5.19  & 1.65  & 99.86 & 1.10  & (132) & 22.57 & 1.01  & (24)  & 5.81  & 0.67  & (3) \\
    \textbf{DBLP Network} & 32.14 & 19.66 & 23.17 & 13.02 & 10.08 & 4.58  & 341.08 & 1.12  & (131) & 64.75 & 1.10  & (24)  & 11.53 & 0.94  & (4) \\
    \textbf{Youtube Network} & 35.19 & 22.86 & 26.80 & 16.53 & 21.18 & 12.90 & 836.64 & 1.13  & (477) & 161.36 & 1.11  & (93)  & 33.08 & 1.05  & (18) \\
    \bottomrule
    \end{tabular}%
\captionof{table}{\label{tbl:dijk-raw}
Number of read and write transfers of different Dijkstra implementations on different cache size.  Numbers are normalized to read or write transfers per vertex.  We ran SSSP queries on 10 different randomly-chosen source nodes.  Numbers in the parentheses are the average phases.}
\bigskip
    \begin{tabular}{l|r@{ }@{ }@{ }r@{ }@{ }@{ }r|r@{ }@{ }@{ }r@{ }@{ }@{ }r||r@{ }@{ }@{ }r@{ }@{ }@{ }r|r@{ }@{ }@{ }r@{ }@{ }@{ }r} \toprule
          & \multicolumn{6}{c||}{$\bm{\wcost{}=10}$}             & \multicolumn{6}{c}{$\bm{\wcost{}=100}$} \\
\cline{2-13}
    & \multicolumn{3}{c|}{\textbf{Classic}} & \multicolumn{3}{c||}{\textbf{Phased Dijkstra}} & \multicolumn{3}{c|}{\textbf{Classic}} & \multicolumn{3}{c}{\textbf{Phased Dijkstra}} \\
\hline
    \multicolumn{1}{c|}{\textbf{Cache Size}} & \multicolumn{1}{c}{\textbf{2k}} & \multicolumn{1}{c}{\textbf{10k}} & \multicolumn{1}{c|}{\textbf{50k}} & \multicolumn{1}{c}{\textbf{2k}} & \multicolumn{1}{c}{\textbf{10k}} & \multicolumn{1}{c||}{\textbf{50k}} & \multicolumn{1}{c}{\textbf{2k}} & \multicolumn{1}{c}{\textbf{10k}} & \multicolumn{1}{c|}{\textbf{50k}} & \multicolumn{1}{c}{\textbf{2k}} & \multicolumn{1}{c}{\textbf{10k}} & \multicolumn{1}{c}{\textbf{50k}} \\
    \hline
    \textbf{2D Grid} & 59.1 & 28.0 & 22.4 & 17.1 & 14.7 & 13.1 & 497.8 & 238.9 & 192.2 & 118.3 & 114.3 & 106.8 \\
    \textbf{3D Grid} & 171.2 & 102.8 & 41.7 & 137.8 & 39.3 & 16.6 & 1478.9 & 871.2 & 343.5 & 239.0 & 140.1 & 113.1 \\
    \textbf{PA Roadmap} & 49.2 & 10.2 & 5.1  & 40.3 & 7.1  & 3.8  & 421.4 & 83.1 & 41.2 & 129.0 & 53.9 & 28.7 \\
    \textbf{TX Roadmap} & 48.7 & 10.4 & 5.1  & 47.6 & 7.1  & 3.7  & 417.5 & 84.3 & 40.9 & 134.5 & 54.5 & 28.3 \\
    \textbf{Stan Webgraph} & {\textit{218.6}} & 147.3 & 49.7 & 414.9 & 91.2 & 22.6 & 1850.2 & 1232.2 & 401.8 & 507.7 & 182.8 & 101.9 \\
    \textbf{NDU Webgraph} & 178.4 & 97.2 & 21.6 & 110.9 & 32.7 & 12.5 & 1559.4 & 836.3 & 169.7 & 209.9 & 123.9 & 72.8 \\
    \textbf{DBLP Network} & {\textit{228.7}} & 153.4 & 55.9 & 352.3 & 75.7 & 20.9 & 1997.6 & 1325.2 & 468.4 & 453.1 & 174.7 & 105.9 \\
    \textbf{Youtube Network} & {\textit{263.8}} & 192.1 & 150.1 & 847.9 & 172.5 & 43.6 & 2321.0 & 1679.4 & 1310.8 & 949.3 & 272.8 & 138.3 \\
    \bottomrule
    \end{tabular}%
\captionof{table}{\label{tbl:dijk-weighted}
The I/O costs on different Dijkstra implementations on different cache size.  The write-read ratio $\wcost{}$ are selected to be typical projected values 10 (latency, bandwidth) and 100 (energy).  Results are based on the numbers in Table~\ref{tbl:dijk-raw}.}
\begin{center}
    \begin{minipage}[t]{0.24\textwidth}
        \includegraphics[width=\textwidth]{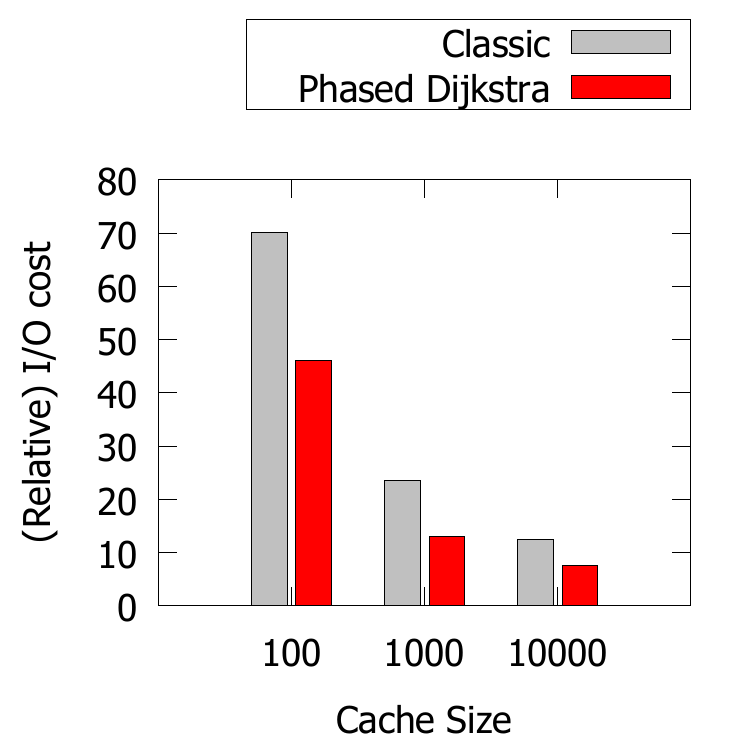}\\ \centering{\footnotesize (a) $\wcost=10$, sparse graphs}
    \end{minipage}
    \begin{minipage}[t]{0.24\textwidth}
        \includegraphics[width=\textwidth]{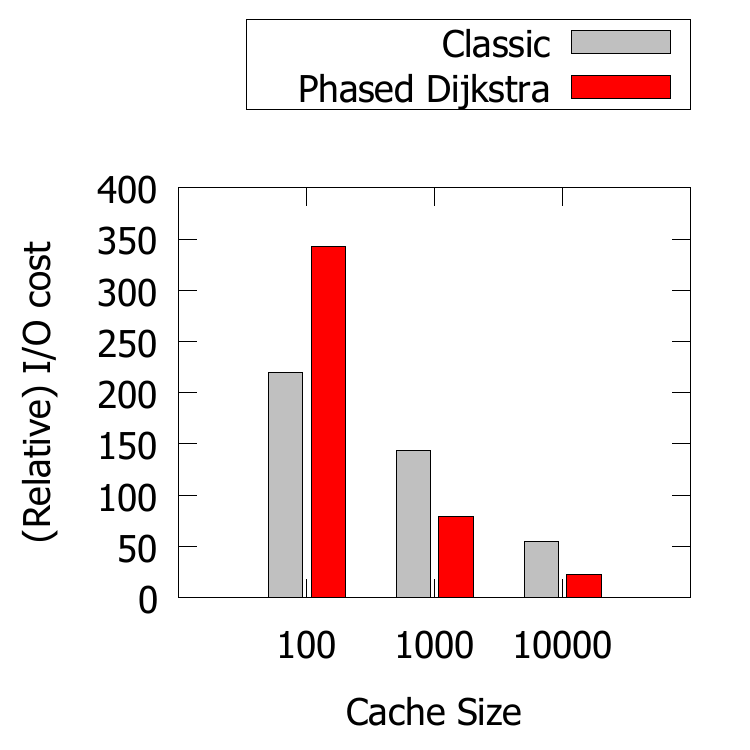}\\ \centering{\footnotesize (b) $\wcost=10$, social networks}
    \end{minipage}
    \begin{minipage}[t]{0.24\textwidth}
        \includegraphics[width=\textwidth]{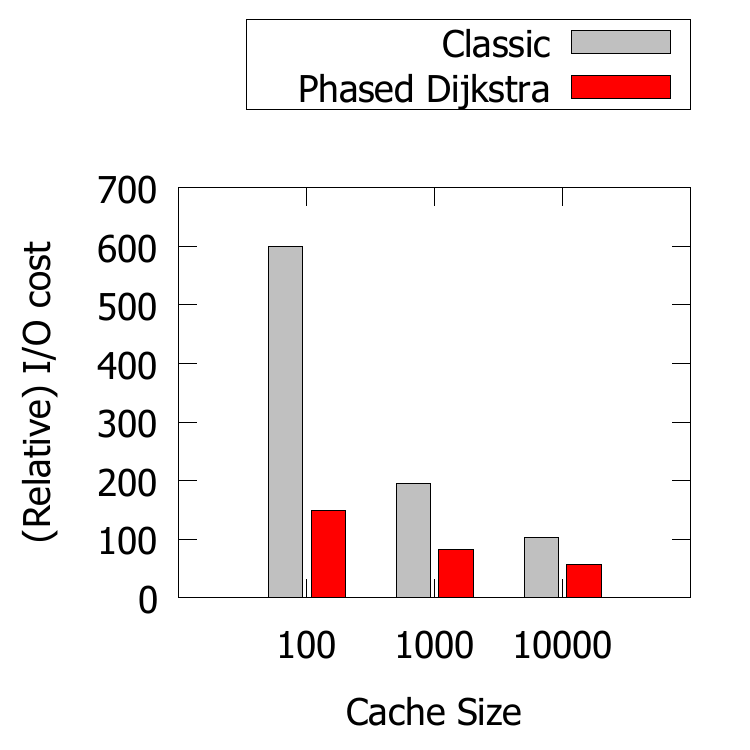}\\ \centering{\footnotesize (c) $\wcost=100$, sparse graphs}
    \end{minipage}
    \begin{minipage}[t]{0.24\textwidth}
        \includegraphics[width=\textwidth]{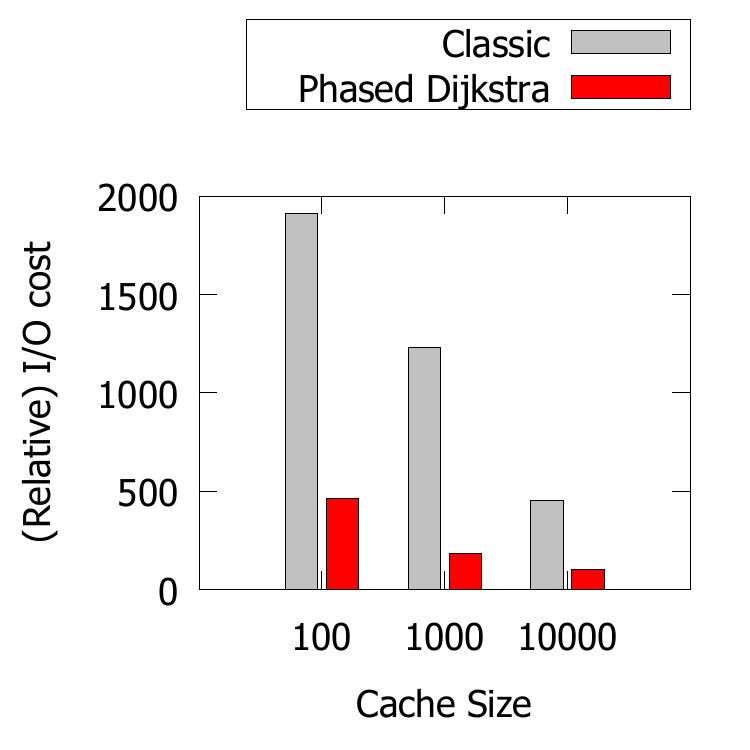}\\ \centering{\footnotesize (d) $\wcost=100$, social networks}
    \end{minipage}
\vspace{.5em}
    \captionof{figure}{\label{fig:dijk} I/O costs of classic Dijkstra and phased Dijkstra.  Graphs are categorized into sparse (almost planar) graphs and social networks and the I/O cost is geometric mean of the four.  Numbers are from Table~\ref{tbl:dijk-weighted}.}
\end{center}
\end{figure*}

We make a special optimization that once all outgoing edges of a vertex are visited, we remove this node and all associated edges in the scan in Line~\ref{line1}.
This is done by using the sign-bit of the output distances, such that it is more likely being cached.
We call the \defn{active set} that contains visited but not removed vertices.

For an efficient implementation that optimizes memory footprint and I/O efficiency, each heap element contains the vertex index, the pointer to the vertex map, and the tentative distance.
In total it takes 16 bytes.
Each element in the vertex map only stores a pointer, and the vertex index can be check via the corresponding heap element.
Throughout our experiment we set the maximum occupancy rate of the 2-level hash table to be $0.8$, and truncation ratio $\epsilon=0.25$.
$M'$ is chosen such that the hash table and priority queue occupy about 40\% of the \smallmem{}, while various values of $M'$ are also discussed.

\subsubsection{Experiments}\label{sec:dijk-exp}

The experiments are run on two Dijkstra implementations with different parameter combinations on cache size, cache policy, and the priority queue size.  
The experiment is run on eight graph instances with various cache size.

Since in phased Dijkstra the number of reads is significantly more than that of writes, to keep the priority queue in the cache, the special cache strategy in Section~\ref{sec:policy} is required.
In Table~\ref{tbl:dijk-policy} we show that different cache policies only cause minor differences, so in the majority of this section we use the \static{} policy.

\myparagraph{Overall performance.}  In Table~\ref{tbl:dijk-raw} we show the number of \rwt{s} of two implementations on different graphs with various cache size.   Cache size varies from 2,000 to 50,000 cache lines each with 64 bytes.  In Table~\ref{tbl:dijk-weighted} the overall I/O costs with different values of $\wcost{}$ are listed based on the numbers in Table~\ref{tbl:dijk-raw}.

For the binary-heap implementation, the actual reads and writes mostly match the theoretical bound $O(m\log (nB/M))$.  Reads are about slightly less than twice as writes: each edge is read once during Dijkstra (linear scan and requires no modification), and an update in the binary heap always requires one more read than writes.
The only exception is on the roadmaps when the frontier size is consistently small while nearby vertices share contiguous indices.
In this case the cache efficiently holds the entire heap and leads to fewer reads and writes to the \largemem{}.

For phased Dijkstra, we first observed that the number of writes is always no more than 1.3 per vertex: one write per vertex when the distance is finalized, plus some other cost to maintain the active set.  The number of \cm{s} is mainly decided the number of phases, and the size of the active set (the edge lists of active vertices at the beginning of each phase needs to be scanned).

The overall I/O performance (shown in Table~\ref{tbl:dijk-weighted}) indicates that phased Dijkstra is consistently better than the binary-heap version except for the only case that both $\wcost{}$ and cache size are small and the active set is large.
This case can hardly happen in practice since the cache size is even much smaller than the current L3 cache.
The improvement on I/O cost in all cases is up to 3 and 7.6 when $\wcost$ is 10 and 100.  This peak occurs when the \smallmem{} sizes is large and frontier size is larger, since at this time the cost to maintain the binary heap in the classic implementation is costly, while the extra cost to run phased Dijkstra is insignificant.

\begin{table}[tp]
  \centering
  \small
  \def\arraystretch{1.1}
    \begin{tabular}{l|rr|rr}
    \toprule
    \multicolumn{1}{c|}{\textbf{Cache Policy}} & \multicolumn{2}{c|}{\textbf\splitpool{}} & \multicolumn{2}{c}{\textbf{\static{}}} \\
\cline{1-5}          & \multicolumn{1}{c}{\textbf{RT}} & \multicolumn{1}{c|}{\textbf{WT}} & \multicolumn{1}{c}{\textbf{RT}} & \multicolumn{1}{c}{\textbf{WT}} \\
    \hline
    \textbf{2D Grid} & 3.74  & 1.11  & 3.62  & 1.11 \\
    \textbf{3D Grid} & 27.43 & 1.12  & 28.16 & 1.12 \\
    \textbf{PA Roadmap} & 1.84  & 0.88  & 1.89  & 0.52 \\
    \textbf{TX Roadmap} & 1.83  & 0.87  & 1.88  & 0.53 \\
    \textbf{Stan Webgraph} & 81.40 & 1.02  & 81.02 & 1.02 \\
    \textbf{NDU Webgraph} & 21.96 & 1.07  & 22.57 & 1.01 \\
    \textbf{DBLP Network} & 65.20 & 1.11  & 64.75 & 1.10 \\
    \textbf{Youtube Network} & 163.38 & 1.12  & 161.36 & 1.11 \\
\bottomrule
    \end{tabular}%
  \caption{\label{tbl:dijk-policy}
  Number of \rwt{} of phased Dijkstra on two different cache policies: the \splitpool{} policy and the \static{} policy.  The cache contains 10,000 cache-lines.
  Different policies give very similar performance except for the writes on roadmaps. }
\end{table}%

\begin{figure*}[!ht]
  \centering
  \small
  \def\arraystretch{1.2}
    \begin{tabular}{l|rr|rr}
    \toprule
    \multicolumn{1}{c|}{\textbf{Cache Policy}} & \multicolumn{2}{c|}{\textbf\splitpool{}} & \multicolumn{2}{c}{\textbf{\static{}}} \\
\cline{1-5}          & \multicolumn{1}{c}{\textbf{RT}} & \multicolumn{1}{c|}{\textbf{WT}} & \multicolumn{1}{c}{\textbf{RT}} & \multicolumn{1}{c}{\textbf{WT}} \\
    \hline
    \textbf{2D Grid} & 3.74  & 1.11  & 3.62  & 1.11 \\
    \textbf{3D Grid} & 27.43 & 1.12  & 28.16 & 1.12 \\
    \textbf{PA Roadmap} & 1.84  & 0.88  & 1.89  & 0.52 \\
    \textbf{TX Roadmap} & 1.83  & 0.87  & 1.88  & 0.53 \\
    \textbf{Stan Webgraph} & 81.40 & 1.02  & 81.02 & 1.02 \\
    \textbf{NDU Webgraph} & 21.96 & 1.07  & 22.57 & 1.01 \\
    \textbf{DBLP Network} & 65.20 & 1.11  & 64.75 & 1.10 \\
    \textbf{Youtube Network} & 163.38 & 1.12  & 161.36 & 1.11 \\
\bottomrule
    \end{tabular}%
  \captionof{table}{\label{tbl:dijk-policy}
  Number of \rwt{} of phased Dijkstra on two different cache policies: the \splitpool{} policy and the \static{} policy.  The cache contains 10,000 cache-lines.
  Different policies give very similar performance except for the writes on roadmaps. }
\bigskip
    \begin{tabular}{l|rrrrrrrr}
    \toprule
   \multicolumn{1}{@{}c@{ }|}{\textbf{Priority Queue Size}}       & \multicolumn{2}{c}{\textbf{30\%}} & \multicolumn{2}{c}{\textbf{40\%}} & \multicolumn{2}{c}{\textbf{50\%}} & \multicolumn{2}{c}{\textbf{60\%}} \\
\cline{1-9}          & \multicolumn{1}{c}{\textbf{RT}} & \multicolumn{1}{c}{\textbf{WT}} & \multicolumn{1}{c}{\textbf{RT}} & \multicolumn{1}{c}{\textbf{WT}} & \multicolumn{1}{c}{\textbf{RT}} & \multicolumn{1}{c}{\textbf{WT}} & \multicolumn{1}{c}{\textbf{RT}} & \multicolumn{1}{c}{\textbf{WT}} \\
    \hline
    \textbf{2D Grid} & 8.4   & 1.10  & 8.6   & 1.11  & 8.9   & 1.11  & 9.3   & 1.11 \\
    \textbf{3D Grid} & 33.7  & 1.12  & 28.9  & 1.12  & 25.6  & 1.12  & 23.3  & 1.12 \\
    \textbf{PA Roadmap} & 2.7   & 0.48  & 2.8   & 0.52  & 3.0   & 0.58  & 3.2   & 0.66 \\
    \textbf{TX Roadmap} & 2.6   & 0.48  & 2.7   & 0.53  & 2.9   & 0.59  & 3.1   & 0.66 \\
    \textbf{Stan Webgraph} & 94.4  & 1.02  & 82.9  & 1.02  & 72.1  & 1.02  & 67.1  & 1.02 \\
    \textbf{NDU Webgraph} & 26.1  & 1.01  & 22.8  & 1.01  & 20.2  & 1.03  & 18.5  & 1.04 \\
    \textbf{DBLP Network} & 75.1  & 1.10  & 64.9  & 1.10  & 57.3  & 1.10  & 51.5  & 1.11 \\
    \textbf{Youtube Network} & 183.8 & 1.11  & 161.5 & 1.11  & 144.4 & 1.12  & 134.9 & 1.12 \\
    \bottomrule
    \end{tabular}%
  \captionof{table}{\label{tab:dijk-perc}Number of read and write transfers of phased Dijkstra on different priority queue's size.  The overall percentage of priority queue and vertex map varies from 30\% to 60\% comparing to the cache size, which is 10,000 cache-lines.}
\bigskip
\end{figure*}%

\myparagraph{Different cache maintenance policy.}
We show the number of \rwt{} of phased Dijkstra on two different cache policies in~\ref{tbl:dijk-policy}.
Different policies actually give a very similar performance in the graph instances.
More analysis is shown in the full version of this paper.

\myparagraph{Picking appropriate priority queue's size $M'$.}  In previous experiment we empirically set the overall size of the priority queue and vertex map to be about 40\% of the overall cache size.  However, this percentage can affect the performance of phased Dijkstra.  Larger percentage leads the heap to contain more elements so that the overall number of phases is decreased.  Meanwhile the size left for the rest cache is smaller, which decreases the cache performance.  Hence, the number of phases and specific graph property lead to different optimality point of the the priority queue's size and no easy conclusions can be drawn.  In Table~\ref{tab:dijk-perc} we show a snapshot on the cache size of 10,000 and the percentage varies from 30\% to 60\%.  The trend is that, when the priority queue's size is smaller than the average frontier size then larger priority queue helps, and vice visa.  In general different priority queue's sizes only make a minor difference on I/O cost and will not affect the relatively lower cost comparing to the binary-heap implementation.

\subsubsection{Conclusions}

We discussed phased Dijkstra and experiment its performance on a variety of graphs.
The high-level idea is to fit the computation within the \smallmem{} (i.e.\ the cache) and thus requires no intermediate writes to the large asymmetry memory.
The experimental results show that phased Dijkstra consistently outperforms the binary-heap version on I/O cost except for the combination of small $\wcost$ ($=10$), small cache size, and on social networks.
Although phased Dijkstra contains some parameters, we also show that they do not affect the efficiency of phased Dijkstra when they are within a reasonable large range.
A similar case also holds for different cache policies.

Notice that the idea here that fits the computation in the \smallmem{} can also be applied to computing minimum spanning tree, sorting, and many other problems.

\newcommand{\etalchar}[1]{$^{#1}$}

\end{document}